\documentclass[aps,prl,nofootinbib,twocolumn,showpacs,superscriptaddress,letterpaper,amsmath,amssymb]{revtex4}

\usepackage{graphicx}  
\usepackage{dcolumn}   
\usepackage{bm}        
\usepackage{amssymb}   
\usepackage{feynmf}
\usepackage{slashed}
\def\gsim{\lower0.5ex\hbox{$\:\buildrel >\over\sim\:$}}
\def\lsim{\lower0.5ex\hbox{$\:\buildrel <\over\sim\:$}}

\newcommand{\be}{\begin{equation}}
\newcommand{\ee}{\end{equation}}
\newcommand{\bea}{\begin{eqnarray}}
\newcommand{\eea}{\end{eqnarray}}

\newcommand{\nbox}{{\,\lower0.9pt\vbox{\hrule \hbox{\vrule height 0.2 cm
\hskip 0.2 cm \vrule height 0.2 cm}\hrule}\,}}

\begin{document}

\thispagestyle{empty}
\vspace*{-3.5cm}

\vspace{0.5in}

\title{Disentangling Instrumental Features of the 130 GeV Fermi Line}

\begin{center}
\begin{abstract}
We  study the instrumental features of photons from the peak observed at $E_\gamma=130$ GeV in the
 spectrum of Fermi-LAT data.  We use the {\sc sPlots} algorithm to reconstruct
 -- seperately for the photons in the peak and
for background photons -- the distributions of incident angles, the
recorded time, features of the spacecraft position, the
zenith angles,  the conversion type and details of the energy and
direction reconstruction. The presence of a striking feature or
cluster in such a variable would suggest an instrumental cause for the
peak.  In the publically available data, we find several suggestive
features which may inform further  studies by instrumental
experts, though the size of the signal sample is too small to draw
statistically significant conclusions.
\end{abstract}
\end{center}

\author{Daniel Whiteson}
\affiliation{Department of Physics and Astronomy, University of California, Irvine, CA 92697}
\maketitle

\section{Introduction}

While the existence of dark matter is widely accepted, its particle
nature remains undiscovered.   Potential avenues for discovery include 
observation of  production at high energy 
accelerators,  scattering with heavy nuclei in large low-noise
underground volumes, or annihilation.

A clear signal of dark matter annihilation may be carried by gamma
rays traveling to Earth from regions in the galaxy of high dark-matter
density.  As they do not typically scatter after their production,
the photon energy and direction are powerful handles for understanding the
mechanism of dark matter annihiliation into standard model particles.

One mechanism is annihilation resulting in quarks, which would
hadronize and yield $\pi^0$ particles which in turn produce
photons. The spectrum of such a process would give fairly low energy
photons ($E_\gamma \le\approx 50$ GeV) which may be
difficult to distinguish from other sources.

A more striking feature may appear from annhilation directly into
two-body final states including a photon. Rather than yielding a broad
energy spectrum,  this process would
produce a photon with a well-defined
energy given (for the process $\chi \chi \rightarrow \gamma Y$) by
\bea
E_\gamma = m_\chi \left( 1 - \frac{M_Y^2}{4 m_\chi^2} \right)
\label{eq:eline}
\eea
where $M_Y$ is the mass of the second annihilation product, such as a
$Z$ boson or a second photon.  For the case 
where $Y=\gamma$, the line occurs at the mass of the dark matter
particle, $E_\gamma=m_\chi$.  This makes a search for peaks in the
photon spectrum an important component of the dark matter 
program using Fermi-LAT data\cite{Abdo:2010nc,Fermi:2012}.

Recent studies have identified a feature in the gamma ray spectrum
near $E_\gamma=130$ GeV\cite{Bringmann:2012vr,Weniger:2012tx} with a source 
close to the galactic center
\cite{Bringmann:2012vr,Weniger:2012tx,Tempel:2012ey,finksu}.  The line
feature is not accompanied by a lower-energy continuum emission, as
would be expected in many models of dark matter
interaction~\cite{wacker}.   However, the large apparant significance
of the feature has generated keen interest in exploring
other, more mundane explanations, such as unconsidered features in the
non-dark-matter background in the difficult region of the galactic
center, or instrumental effects in the Fermi-LAT detector.

In this paper, we present a first study of the instrumental characteristics of photons in the line feature,
using the {\sc sPlots}~\cite{splots} algorithm to disentangle the two
populations (background and peak). This allows us to reconstruct distributions in
variables which may reveal instrumental issues that would not
otherwise be apparant.

\section{sPlots}

In a sample of events with multiple sources, if one variable can be
used to discriminate between the sources, the {\sc sPlots}
algorithm can reconstruct the statistical distribution of each of
the sources in other variables, which we refer to as the `unfolding variables'. {\sc sPlots} uses only information from the discriminating variable
and knowledge of the probability density functions (pdf) for each source in
the discriminating variable.  In addition, the algorithm assumes that
the pdfs can be factorized between the discriminating and unfolding variables.

For the purposes of clarity, we simplify the general {\sc sPlots} formalism
of Ref.~\cite{splots} into the two-sources case we will apply to the
Fermi-LAT data.

Given  pdfs for two sources $f_1(y)$, and $f_2(y)$ in the discriminating variable
$y$, one can construct a histogram in another unfolding variable $x$ using
weights for each source class, $sP_1$ and $sP_2$, defined as:

\[ sP_1(y) = \frac{\textbf{V}_{11}f_1(y) + \textbf{V}_{12}f_2(y)}{N_1 f_1(y) + N_2 f_2(y)} \]
\[ sP_2(y) = \frac{\textbf{V}_{21}f_1(y) + \textbf{V}_{22}f_2(y)}{N_1 f_1(y) + N_2 f_2(y)} \]

\noindent
where $N_1$ and $N_2$ are the number of events in each class, as
extracted by a likelihood fit of $f_1$ and $f_2$ to the observed
distribution in $y$, and the inverse of the matrix $\textbf{V}$ is a symmetric $2\times2$ matrix defined as

\[ \textbf{V}^{-1}_{ab} = \sum_{i=1}^{N}  \frac{(N_1+N_2)f_a(y_i)f_b(y_i)}{(N_1 f_1(y_i) + N_2 f_2(y_i))^2} \]

A histogram $h$ in the unfolding variable $x$ can then be constructed for
source 1 as

\[ h_i = \sum^{N_i}_{j=1} sP_1(y_{ji}) \]

\noindent where $i$ is the bin index in the $x$ variable, $N_i$ is the
number of events in that bin, and $y_{ji}$ is the value of the $y$
variable for the $j$th event in the $i$th bin. A histogram for source 2 would be
constructed by replacing $sP_1 \rightarrow sP_2$.

This technique is superior to simply making a selection in the $y$
variable to enhance the relative contribution of one source, which
may still be significantly polluted by the other source. Note that if the two sources were completely separable in the $y$ variable, then
the sPlot weights would reduce trivially to 0 or $1/N$.   This appears to be the first
application of this algorithm to an astrophysical problem~\cite{astrocite}.

\subsection{Example of {\sc sPlots} in Toy Data}

As an illustrative example, we generate toy data from two sources according to the
pdfs:

\[ f_{\textrm{peak}}(x,y) = \frac{1}{\sqrt{2\pi}}e^{-\frac{1}{2}(y-5)^2} \times
\frac{10-x}{50} \]

and 

\[ f_{\textrm{non-peak}}(x,y) = \frac{x}{50} \]

\noindent
where $ f_{\textrm{peak}}(x,y)$ is normally distributed in $y$ with $\sigma=1$, while
$f_{\textrm{non-peak}}(x,y)$ is uniform in $y$, providing good
discrimination power.  Our goal is to construct histograms which
reveal the distribution in $x$ for each of the two sources.

\begin{figure}[h]
\begin{picture}(40,40)
\put(0,0){\includegraphics[width=1.68in]{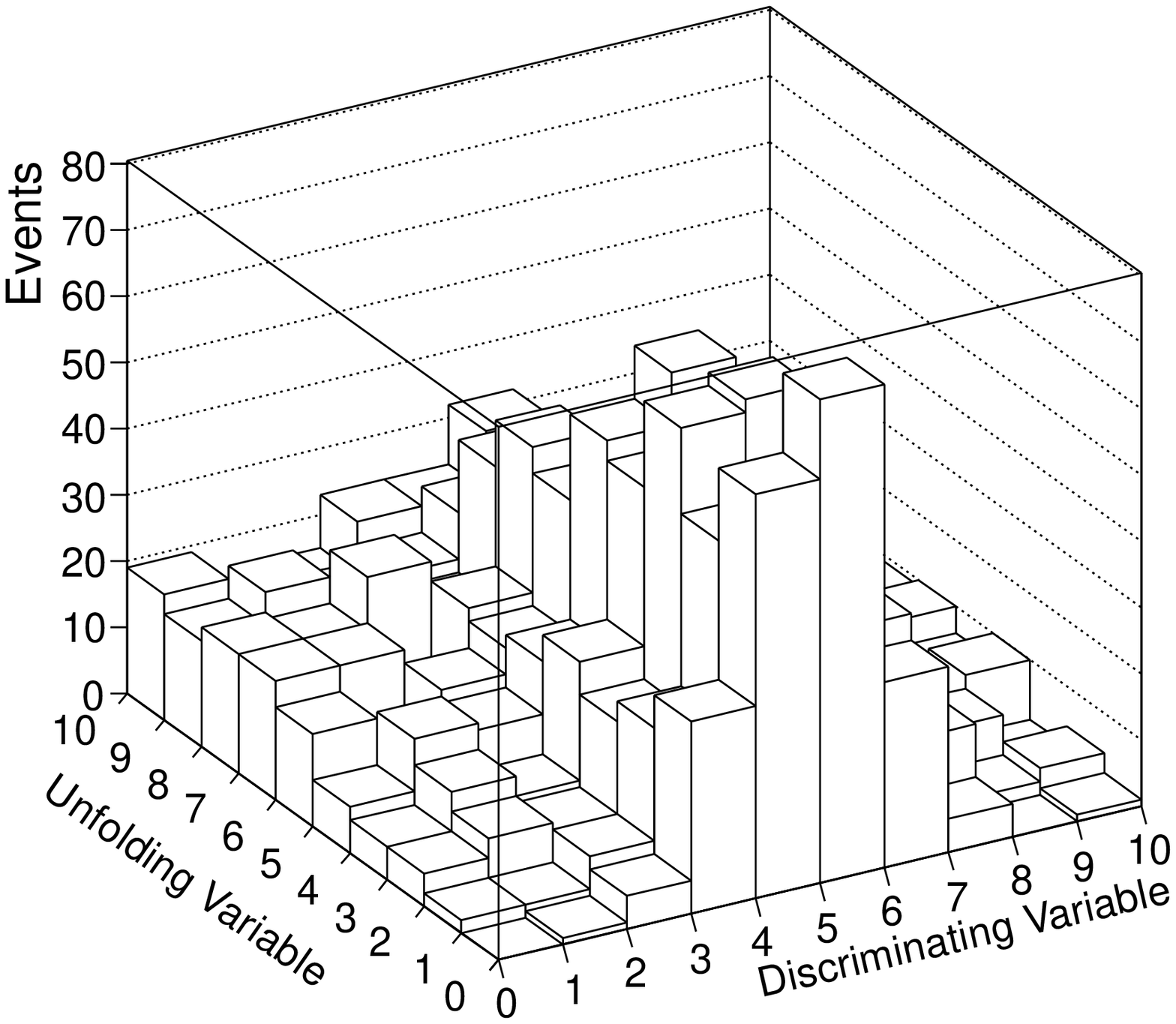}}
\put(8,35){(a)}
\end{picture}
\begin{picture}(40,40)
\put(0,0){\includegraphics[width=1.68in]{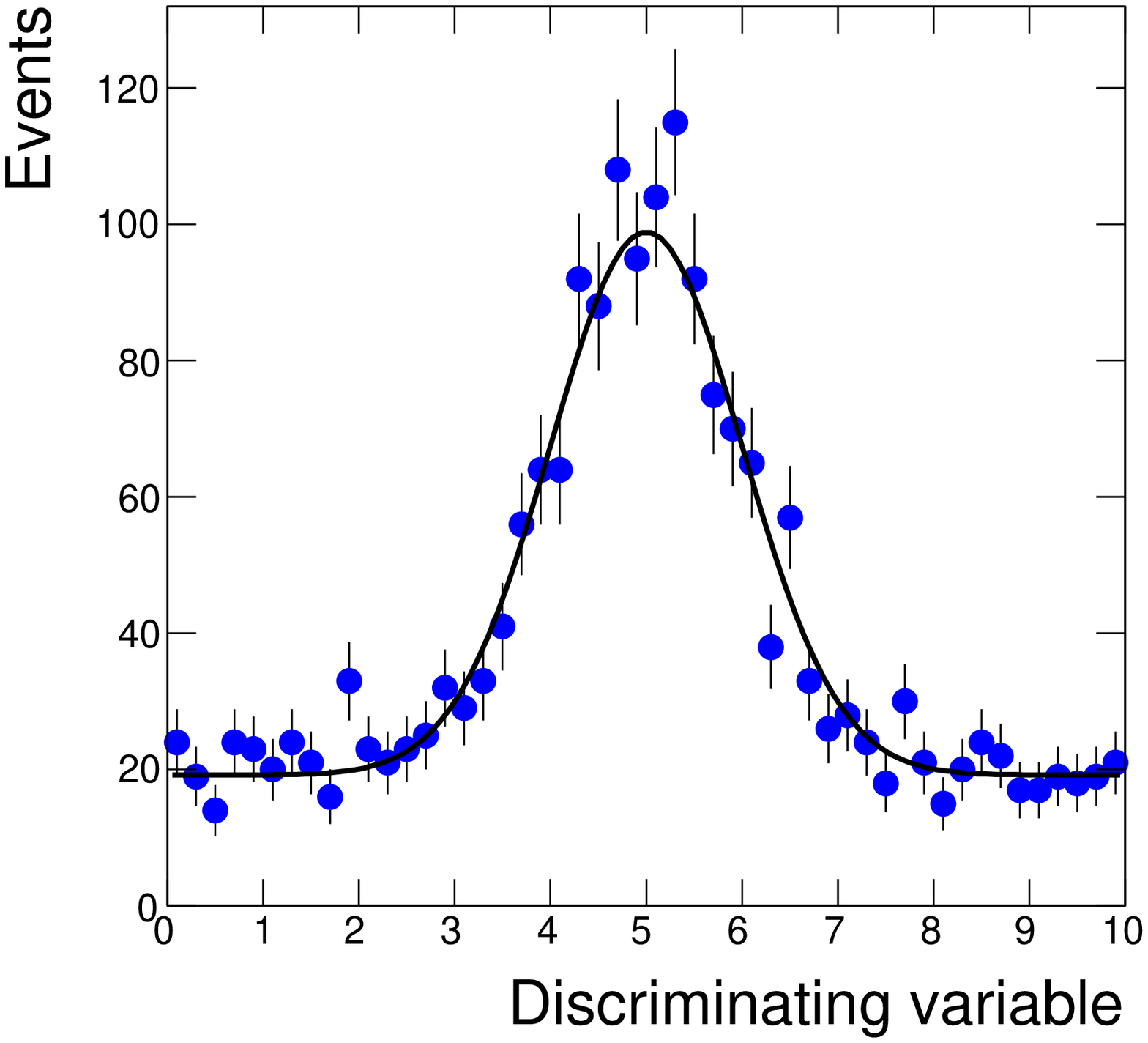}}
\put(8,32){(b)}
\end{picture}
\begin{picture}(90,40)
\put(0,0){\includegraphics[width=3.45in]{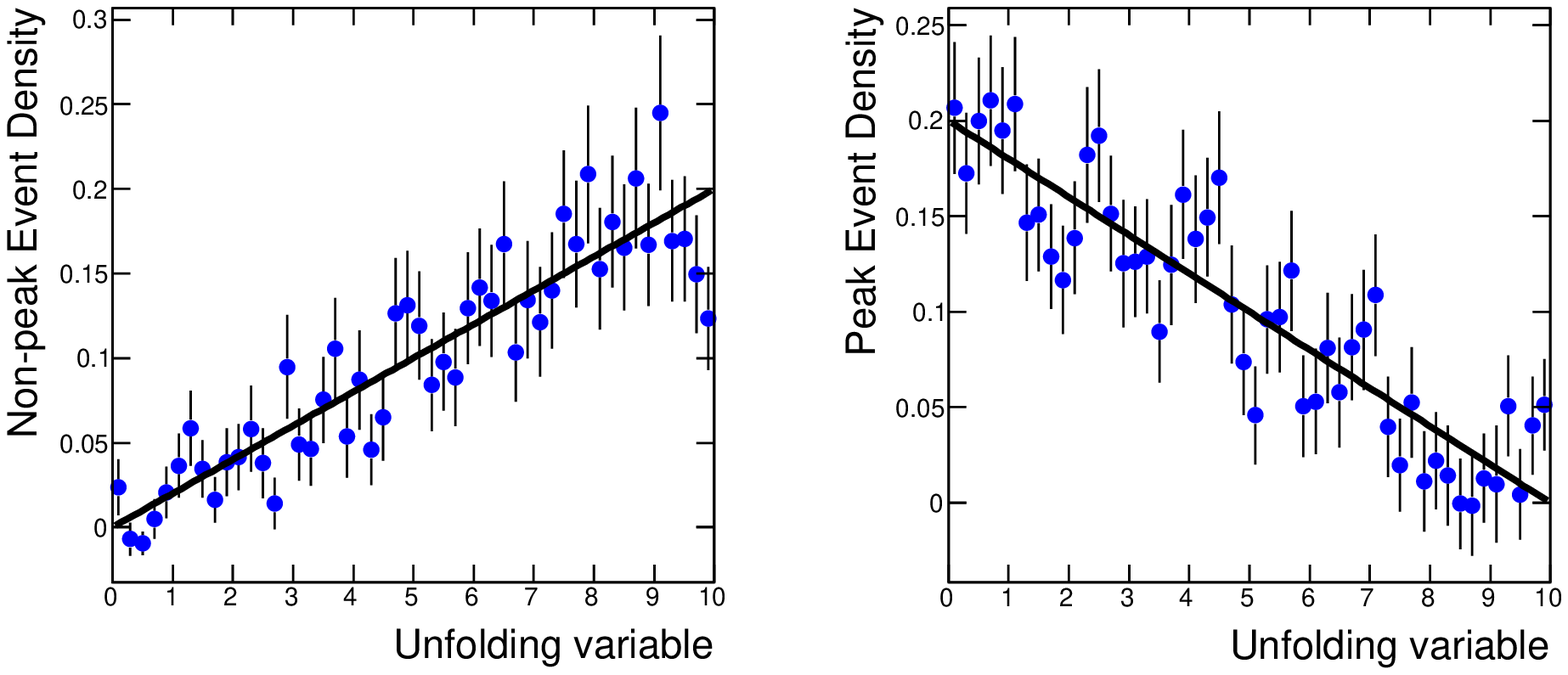}}
\put(9,32){(c)}
\put(75,32){(d)}
\end{picture}
\caption{Example application of the {\sc sPlots}~\cite{splots} algorithm in toy data. Top left, distribution of
  toy data in the discriminating and unfolding variable. Top
  right, distribution of peak and non-peak events in the discriminating
  variable, with the fitted combined pdf. Bottom left (right) shows
  the non-peak (peak) distribution in the unfolding variable (blue points), with the
  true pdf (black line). The unfolded distributions are calculated by {\sc sPlots} using
  only information from the discriminating variable.}
\label{fig:ex}
\end{figure}

Figure~\ref{fig:ex}(a) shows the generated
event distribution in the discriminating variable $y$ and the unfolding variable $x$
using 1000 events from each source. Figure~\ref{fig:ex}(b) shows the
projection in $y$ and the result of the fit to extract $N_1$ and $N_2$, which uses only this
one-dimensional projection and the $y$-dependence of the pdfs.

The unfolded distributions for each source are shown in
Figure~\ref{fig:ex}(c,d), along with the true  pdfs in $x$, which were
not used in reconstructing the unfolded distributions.  The {\sc sPlots}
algorithm is successfully able to disentangle the two sources and reveal the
$x$-dependence of each.  

Note that since the distributions use a
statistical unfolding (rather than event-by-event) and is unaware of
physical constraints, it is possible to have a negative prediction in
a bin.  The statistical uncertainty in a given bin is
$\Delta N = \sqrt{\sum_{i}sP^2}$.  

\begin{figure}[h]
\begin{picture}(90,40)
\put(0,0){\includegraphics[width=3.45in]{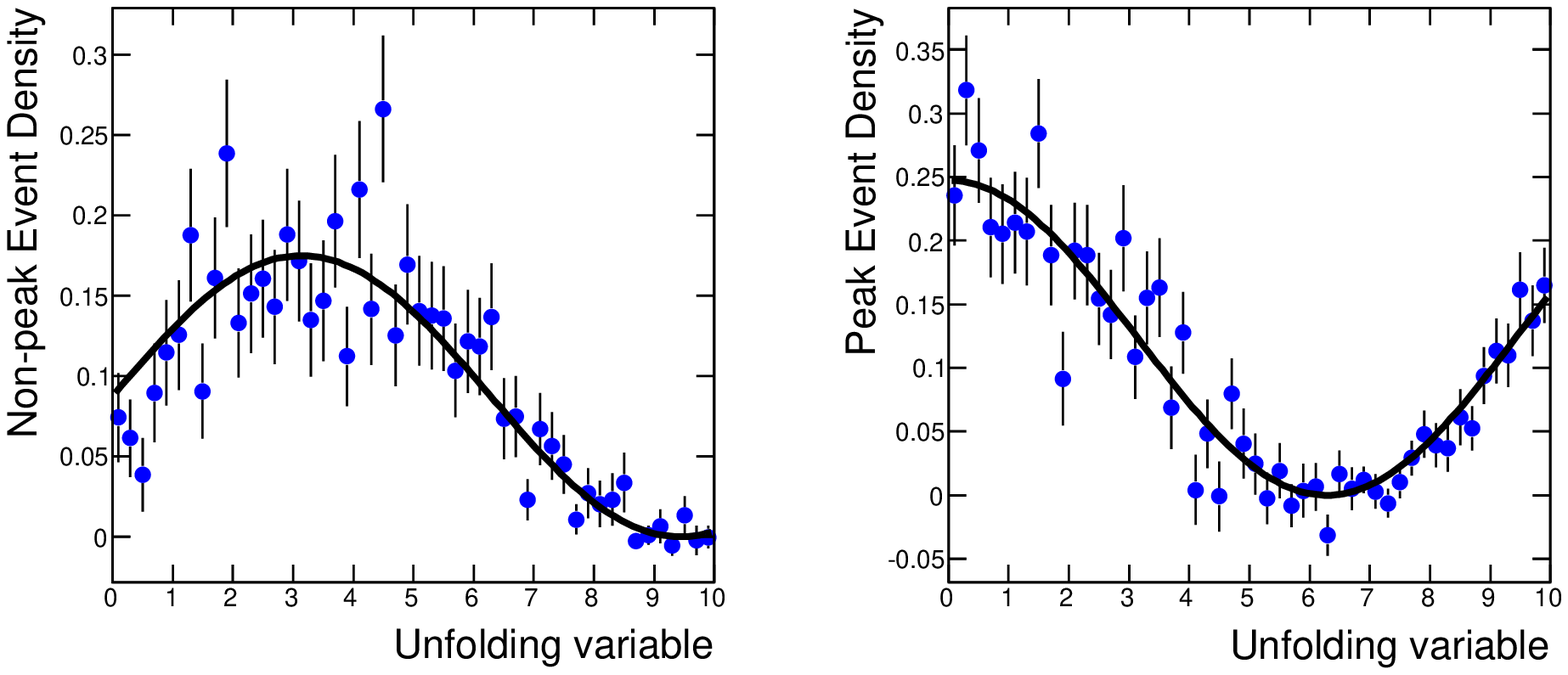}}
\put(9,32){(a)}
\put(75,32){(b)}
\end{picture}
\begin{picture}(90,40)
\put(0,0){\includegraphics[width=3.45in]{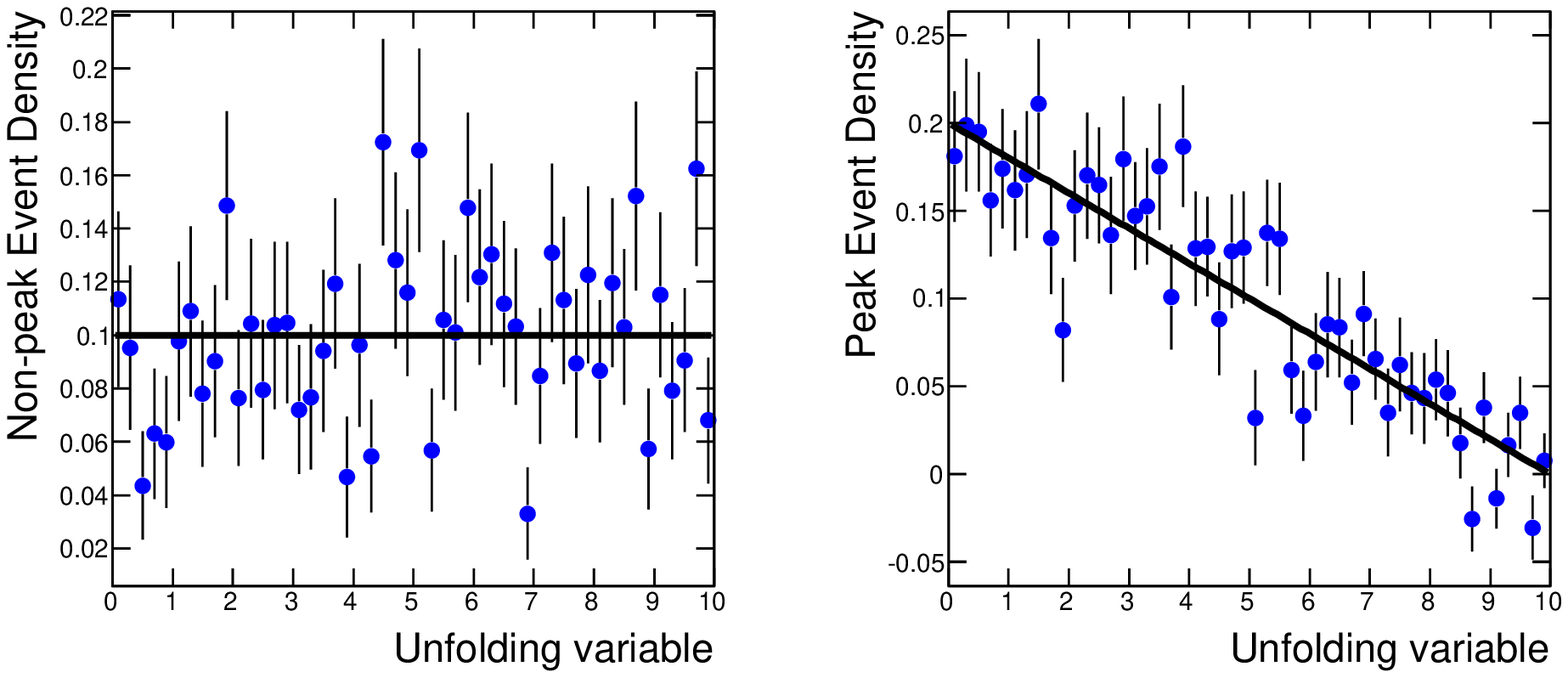}}
\put(9,32){(c)}
\put(75,32){(d)}
\end{picture}
\begin{picture}(90,40)
\put(0,0){\includegraphics[width=3.45in]{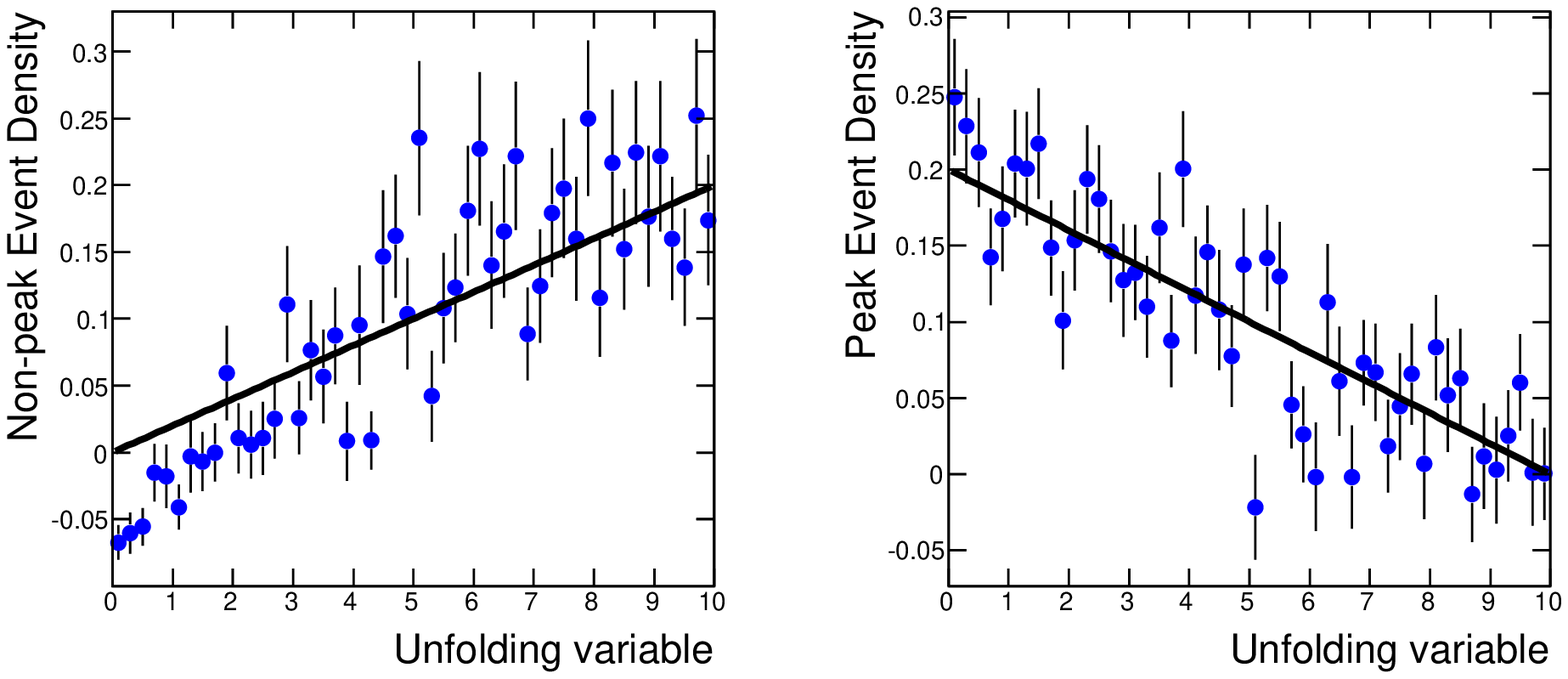}}
\put(9,32){(e)}
\put(75,32){(f)}
\end{picture}
\caption{ Tests of the robustness of {\sc sPlots}. In each case, blue
  points show the {\sc sPlots} reconstructed distribution and the
  black line is the true pdf. Top, reconstruction
  of the unfolding variable with non-linear pdfs.  Center and bottom,
  impact of correlations in the pdfs on the reconstruction of the
  unfolding variable. Center uses a slope which varies with the
  discriminating variable, averaging to zero. Bottom uses a variable-width
  peak. See text for details.}
\label{fig:cor}
\end{figure}

The unfolded distributions can be reconstructed just as well for
non-linear pdfs, see the example in Fig.~\ref{fig:cor}(a,b), where
trigonometric functions have replaced the $x$ dependence of the pdfs.

Correlations between $x$ and $y$ in the pdfs can lead to biases, but
do not catastrophically undermine the unfolding.  For example, if we
use

\[ f_{\textrm{non-peak}}(x,y) = \frac{y+ (1-y/5)x}{50} \] 

so that the slope in $x$ varies from positive to negative over $y\in[0,10]$:

\[ f_{\textrm{non-peak}}(x,0) = \frac{x}{50},
f_{\textrm{non-peak}}(x,10) = \frac{10-x}{50}, \] 

\noindent
then sPlots recovers the average $x$ dependence for the non-peak
source, see Fig.~\ref{fig:cor}(c,d), and the correct dependence for the
events from the peak.

If instead the width of the peak in $y$ varies with $x$, there may be
some biases introduced, but the effects are minor, even for a doubling of the peak width over the range $y\in[0,10]$, see
Fig~\ref{fig:cor}(e,f).

Note that these types
of correlations would be even more troublesome for the traditional
strategy of making a selection to enhance or suppress one source.

\section{The Fermi-LAT data sample}

We use the publically available data with the extended photon data from the Fermi-LAT collaboration
through June 28th, 2012,
making  standard quality requirements~\cite{qual} and examining a
square region around the galactic center, with galactic longitude $-5<l<5$
degrees and galactic latitude $-5<b<5$ with energy greater $E_{\gamma}\ge 50$
GeV.

Other than the reconstructed energy, the photons have other measured
characteristics~\cite{fermidefs} which may give insight into instrumental effects:
\begin{itemize}
\item incident angle $\theta$, measured with respect to the top-face normal of
  the LAT,
\item  azimuth angle $\phi$, measured with respect to the top-face
  normal of
  the LAT, folded as described in Eq. (15) of Ref.~\cite{blah:2012kca}.
\item zenith angle, measured with respect to the zenith line,
  which passed through the earth and LAT's center of mass,
\item earth azimuth angle, the azimuthal angle relative to the same
  line as the zenith, defined such that zero indicates the photon came
  from the northern direction,
\item mission elapsed time, measured relative to January 1, 2001,
\item conversion type (front or back), indicates whether the event
  induced pair production in the front (thin) layers or the back
  (thick) layers of the tracker,
\item the probability that the best energy chosen from the three energy estimators is correct,
\item the probability that the direction estimate is good,
\item ratio of true/raw energy,
\item first layer of the tracker with a hit,
\item the magnetic field in which the LAT is immersed, as
  parameterized by the McIlwain $B$ and $L$
  parameters~\cite{mcilwain}, 
\item the distance from the center of the South Atlantic anomaly,
  calculated as $\sqrt{\Delta \textrm{long}^2 + \Delta
    \textrm{lat}^2}$ in terms of Earth latitude and longitude, and
\item the geomagnetic latitude of the spacecraft.
\end{itemize}

In this paper,  we study the distribution in these variables for signal-like and
background-like photons.   In some cases, a large difference in the
distribution of signal-like and background-like photons would be a
clear indication of an instrumental issue. This is especially true for
variables related to the spacecraft position, environment or angle
(mission time, magnetic field, earth azimuth angle, distance from the
SA anomaly, geomagnetic latitude). Other variables are connected to the quality or class of the
reconstruction (incident angles, conversion type, reconstruction
details) and would give more subtle clues as to whether the feature is
due to a sub-class of photons, or photons with particularly high or
poor resolution.   The response of the LAT is dependent on some of these
variables. For example, the energy resolution is a function of the incident angle
$\theta$ and the conversion type, see Fig.~\ref{fig:edisp}. In
addition, the point-spread function depends on the point in the LAT
where the photon converts.

These indications would be only the
first clues, and would need detailed follow-up by the instrument
experts; a complete study is not possible in the information available
in the public data. The Fermi-LAT collaboration has already performed
detailed studies of the instrument performance and calibration,
including studies of potential systematic biases~\cite{Atwood:2009ez,Abdo:2009gy,blah:2012kca}.

\subsection{Single-Line Analysis}

To analyze the features of the Fermi-LAT data using {\sc sPlots}, we must
define background and signal pdfs in the discriminating variable,
$E_\gamma$.  The background pdf is a simple power-law:

\[ f_{\textrm{bg}}(E_\gamma|\beta,\alpha) = \beta\left(
  \frac{E_\gamma}{E_0}\right) ^{-\alpha} \]

\begin{figure}
\includegraphics[width=3in]{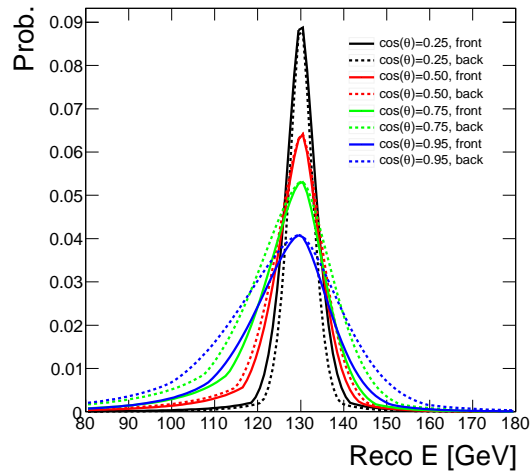}
\caption{Probability density function in Fermi-LAT reconstructed photon energy for photons with
  true energy of $E_\gamma=130$ GeV, for varying choices of the
  incident angle $\theta$ and the conversion type~\cite{fermiedisp}.}
\label{fig:edisp}
\end{figure}

\begin{figure}
\includegraphics[width=3in]{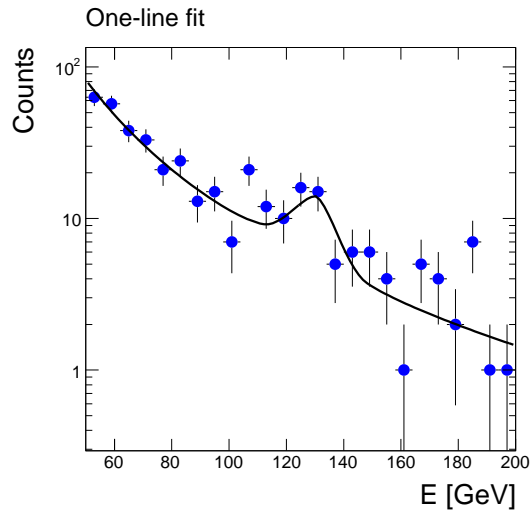}
\caption{Energy of Fermi-LAT photons with signal plus background fit, using a
  single-line hypothesis at $E_\gamma=130$ GeV.}
\label{fig:fit1}
\end{figure}

For the  observed feature we assume a single line (the two-line
hypothesis is discussed below) where the pdf $f_{\textrm{line}}(E_\gamma|E_{\textrm{line}})$ is defined according to the
Fermi-LAT energy dispersion tools definition~\cite{fermiedisp} with a
true photon energy of $E_{\textrm{line}}$, see Fig.~\ref{fig:edisp}.
Applying these pdfs to the observed photon energy spectrum yields the fit seen in Fig.~\ref{fig:fit1}.

\begin{figure}
\includegraphics[width=1.6in]{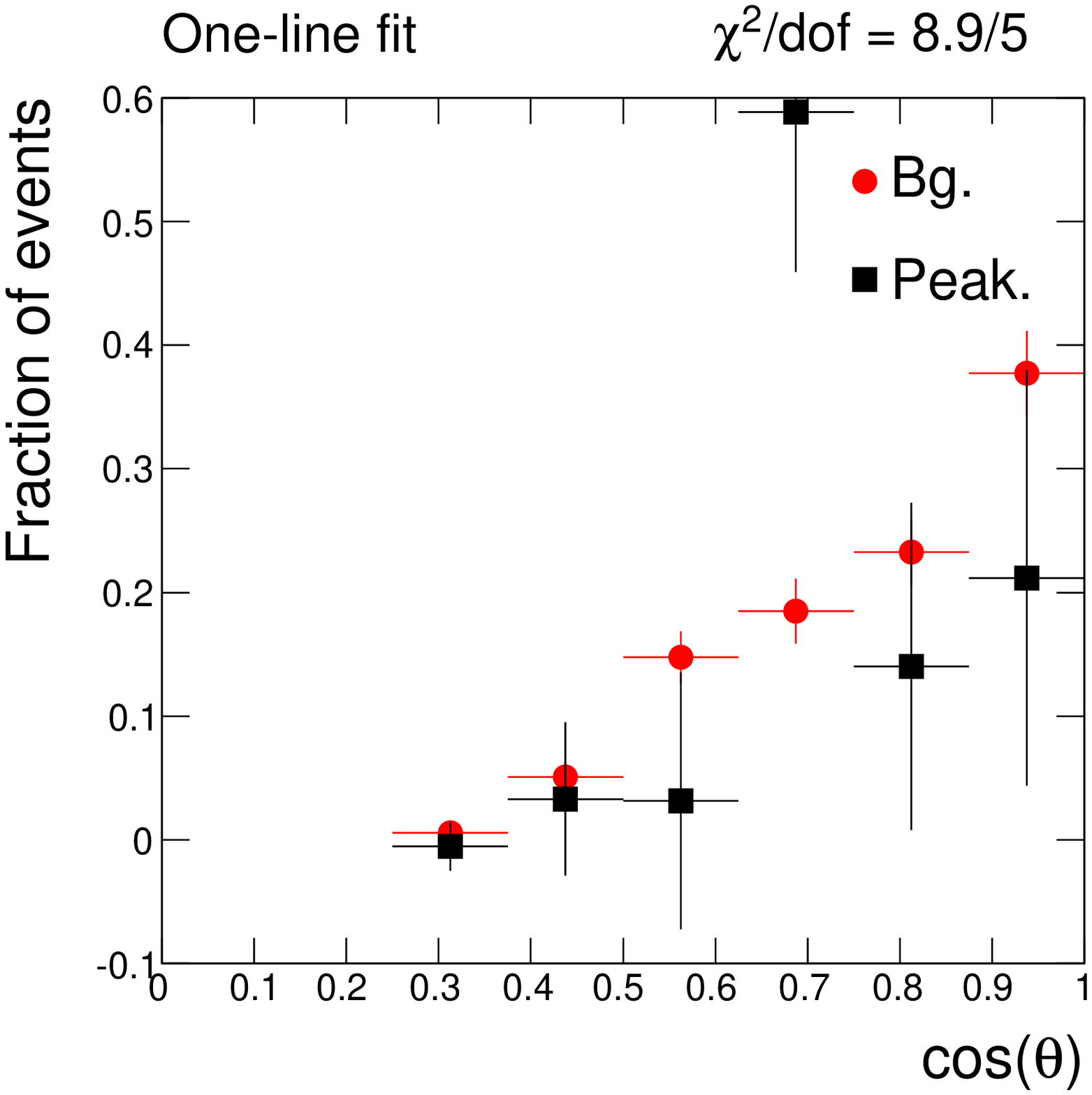}
\includegraphics[width=1.6in]{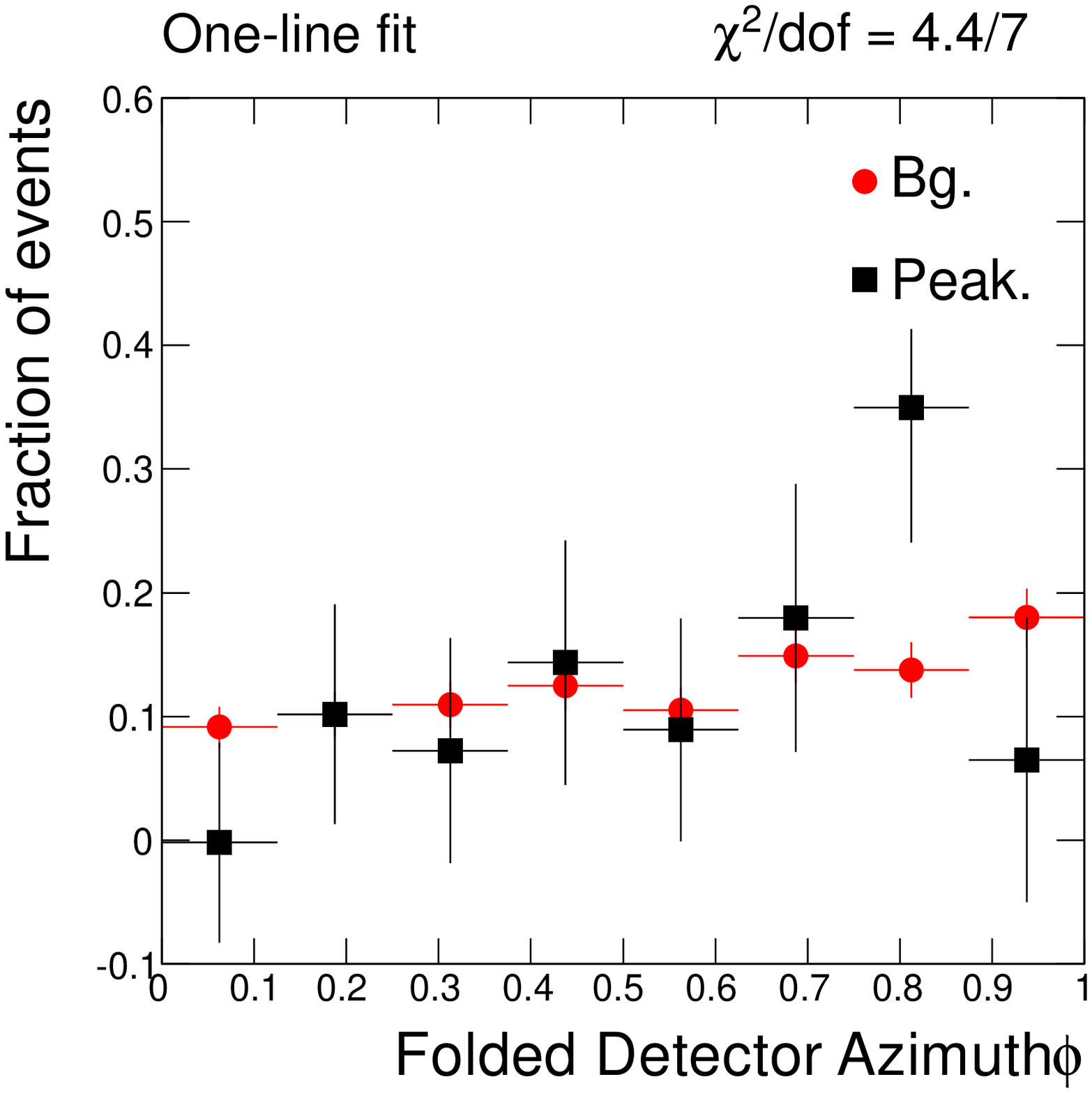}
\caption{Disentangled signal and background distributions. Left,  $\cos(\theta)$ where
  $\theta$ is the photon incidence angle relative  to a line normal
  the Fermi-LAT face. Right, $\phi$, the photon incidence angle
  relative to the sun-facing side~\cite{fermidefs}.}
\label{fig:detang1}
\end{figure}

\begin{figure}
\includegraphics[width=1.6in]{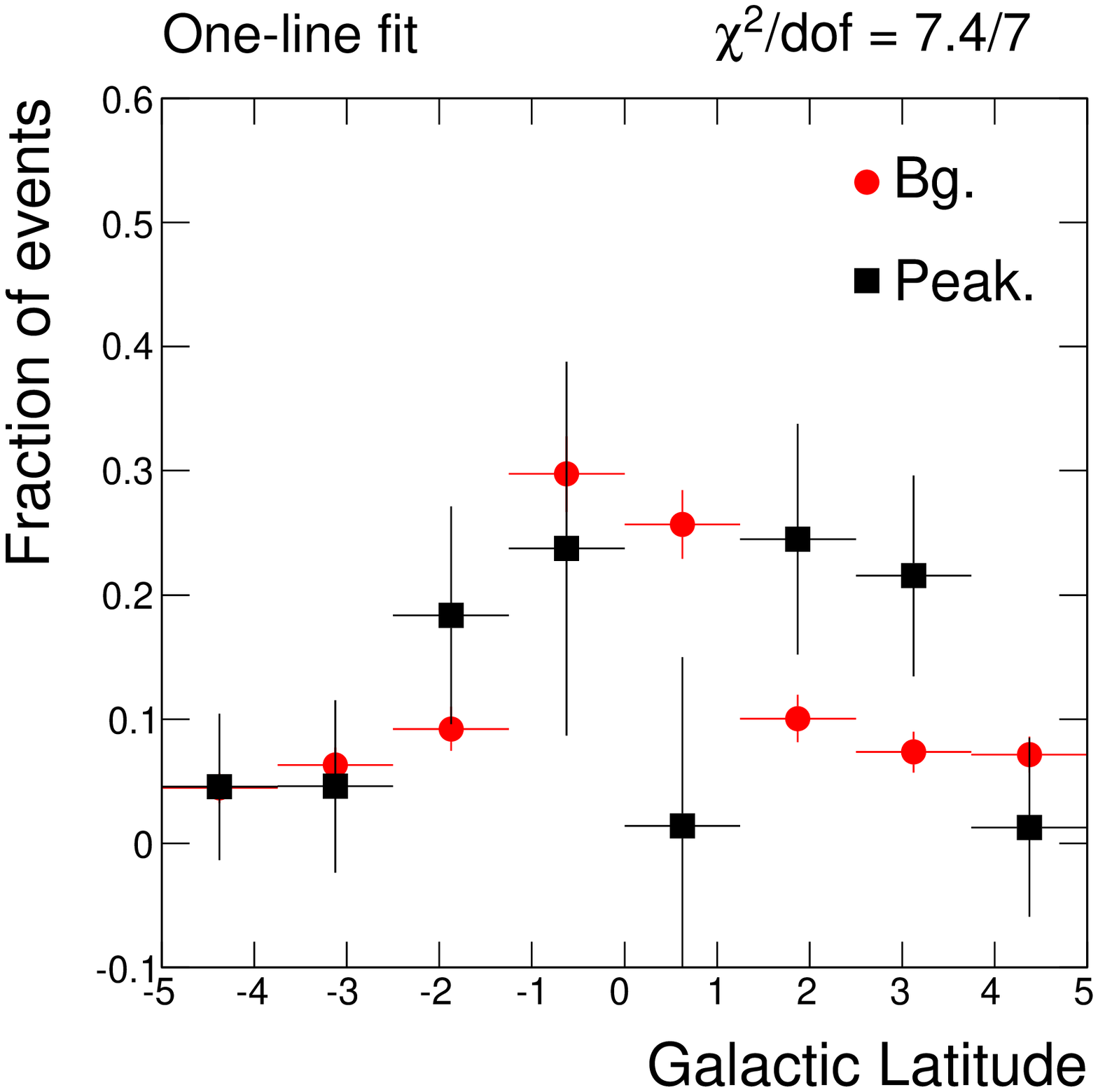}
\includegraphics[width=1.6in]{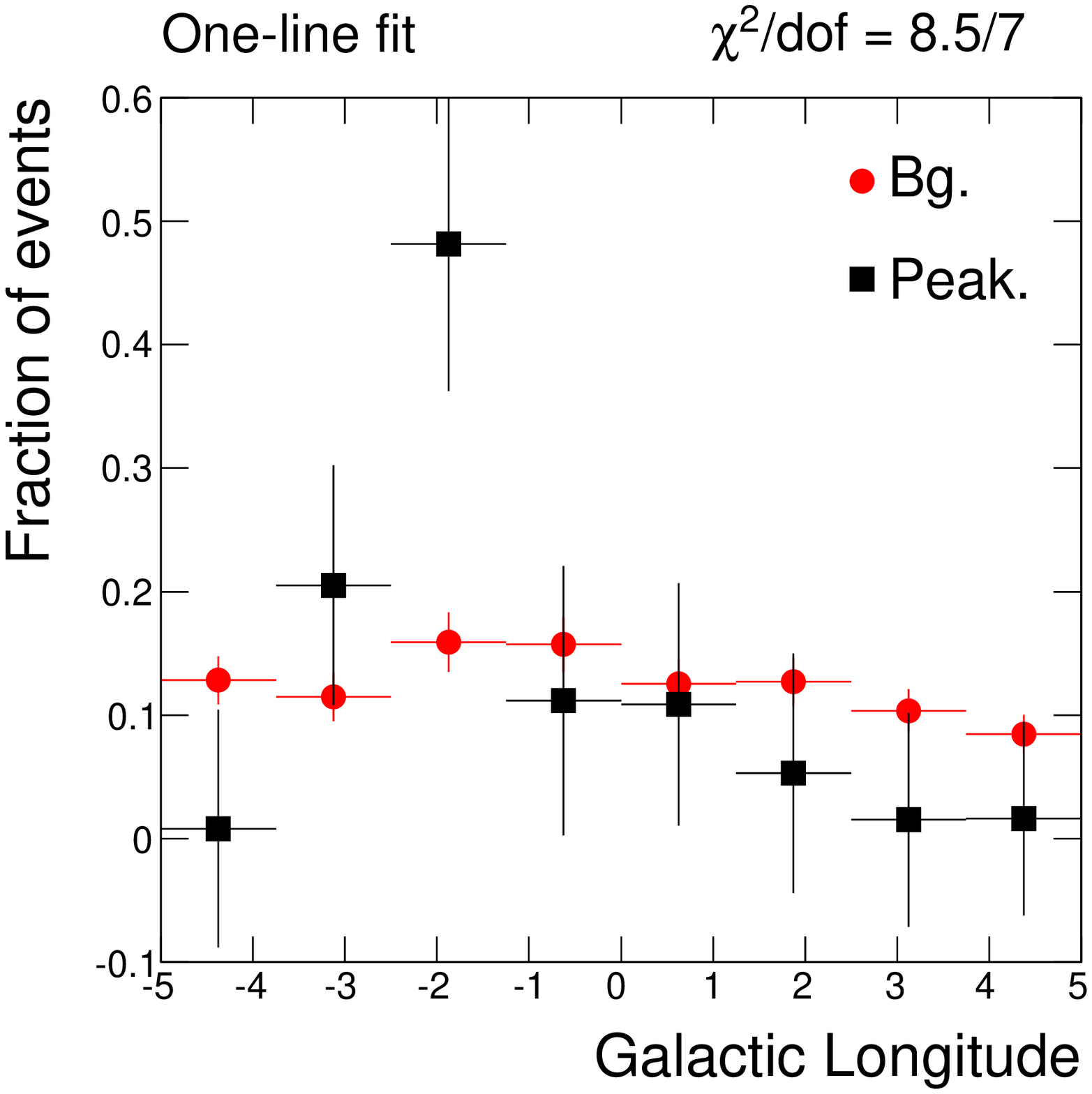}
\caption{Disentangled signal and background distributions in galactic
  coordinates, $b$ (latitude) and $l$
  (longitude)~\cite{fermidefs}. No smoothing has been applied.}
\label{fig:coord1}
\end{figure}

\begin{figure}
\includegraphics[width=1.6in]{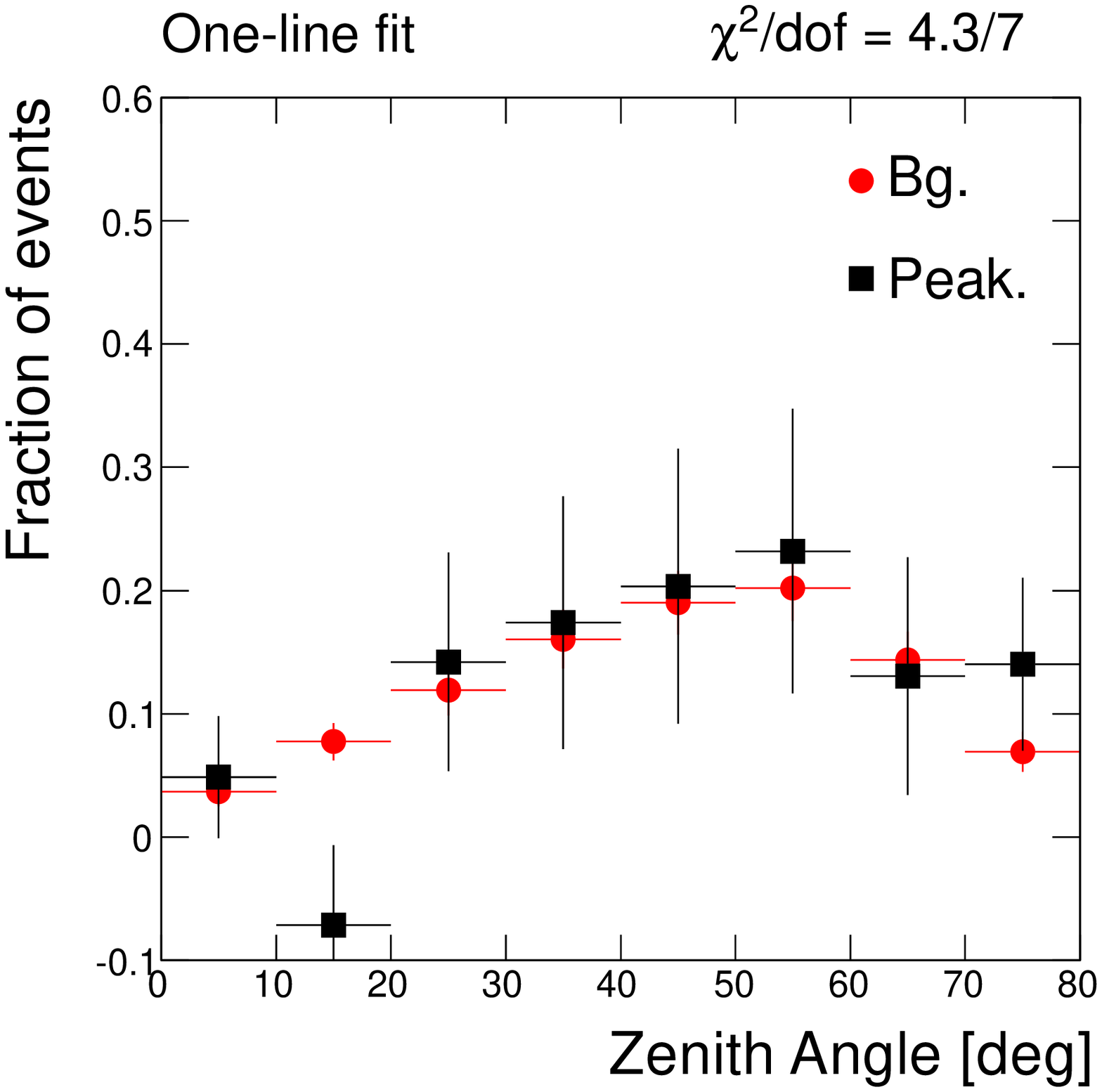}
\includegraphics[width=1.6in]{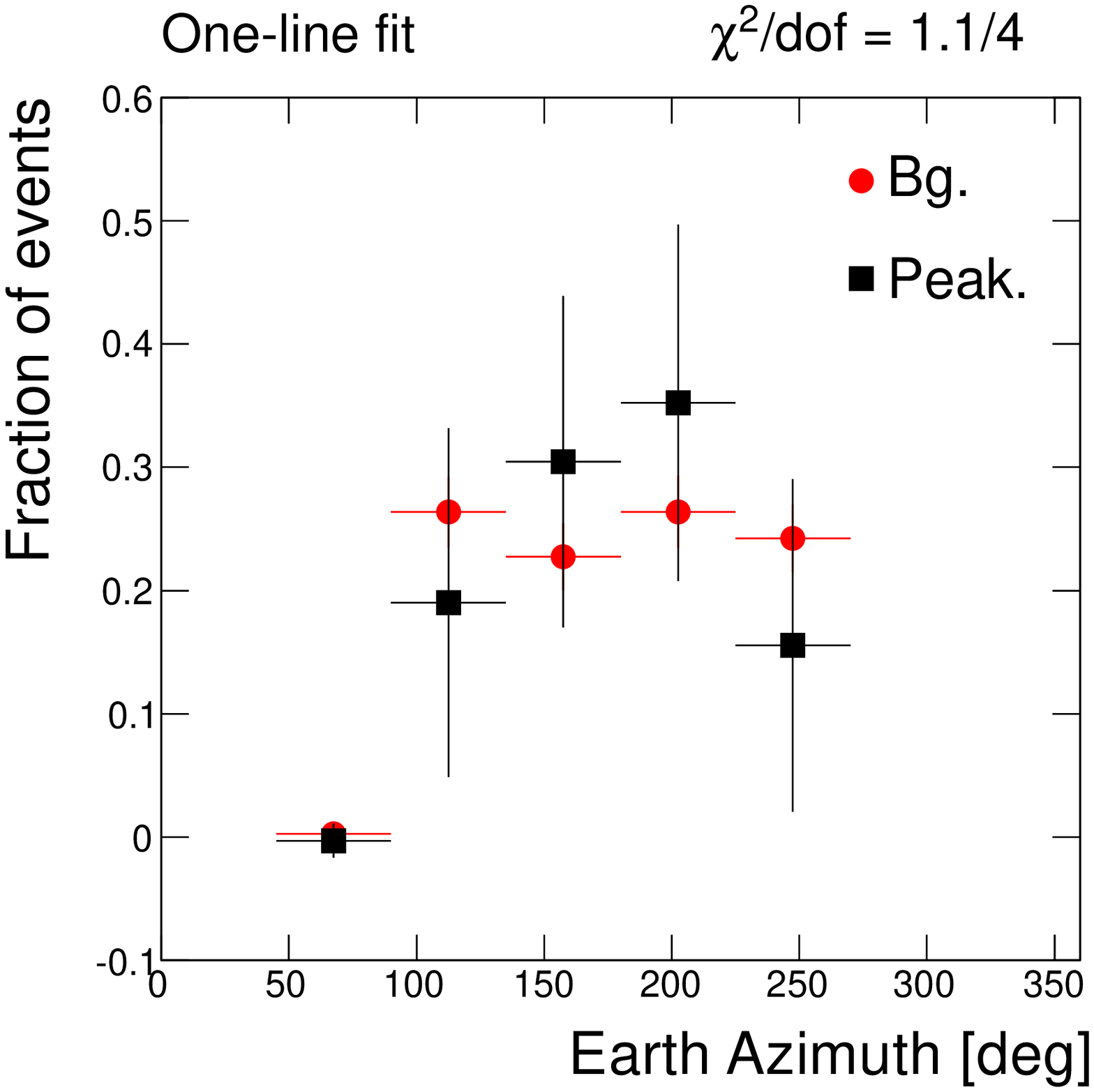}
\caption{Disentangled signal and background distributions. Left, angle
  between the reconstructed photon direction and the zenith line,
  which passed through the earth and Fermi's center of mass. Right,
  the earth azimuth angle. }
\label{fig:zenith1}
\end{figure}

\begin{figure}
\includegraphics[width=1.6in]{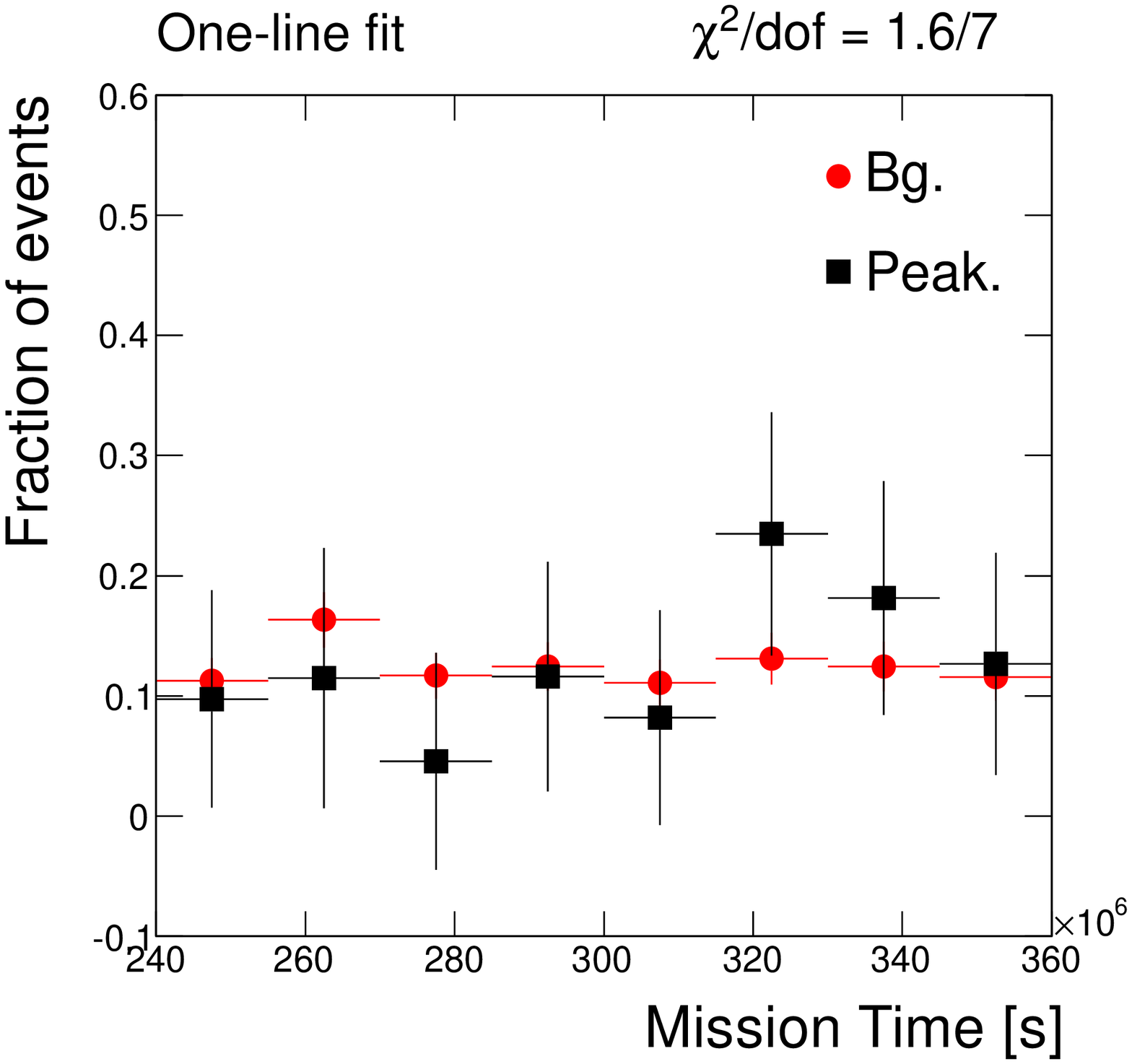}
\includegraphics[width=1.6in]{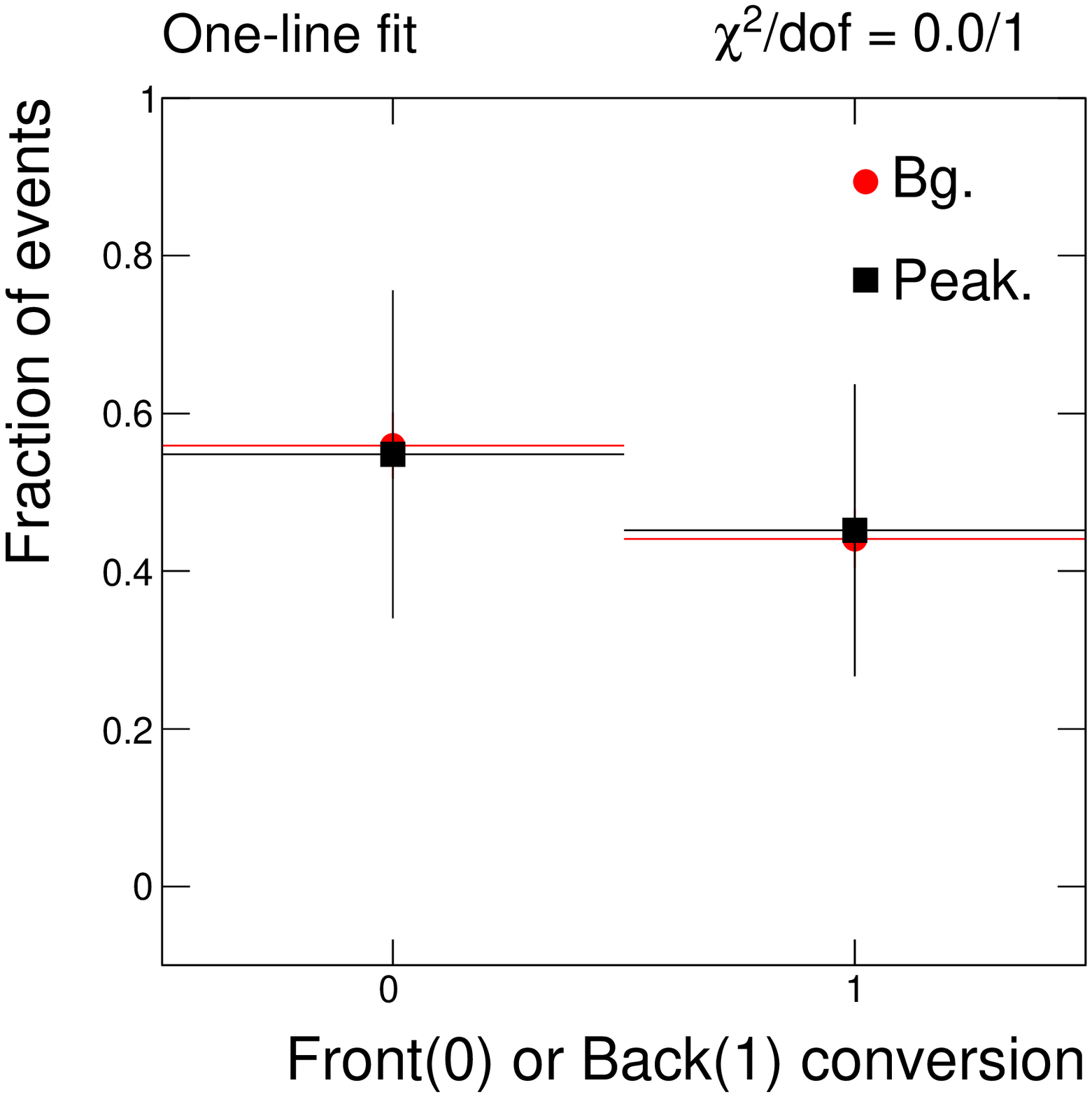}
\caption{Disentangled signal and background distributions.  Left, the mission elapsed time since Jan
  1 2001~\cite{fermidefs}. Right, 
  fraction of events in which the pair production is induced in the
  front (thin) or back (thick) layers of the tracker.}
\label{fig:timeback1}
\end{figure}

\begin{figure}
\includegraphics[width=1.6in]{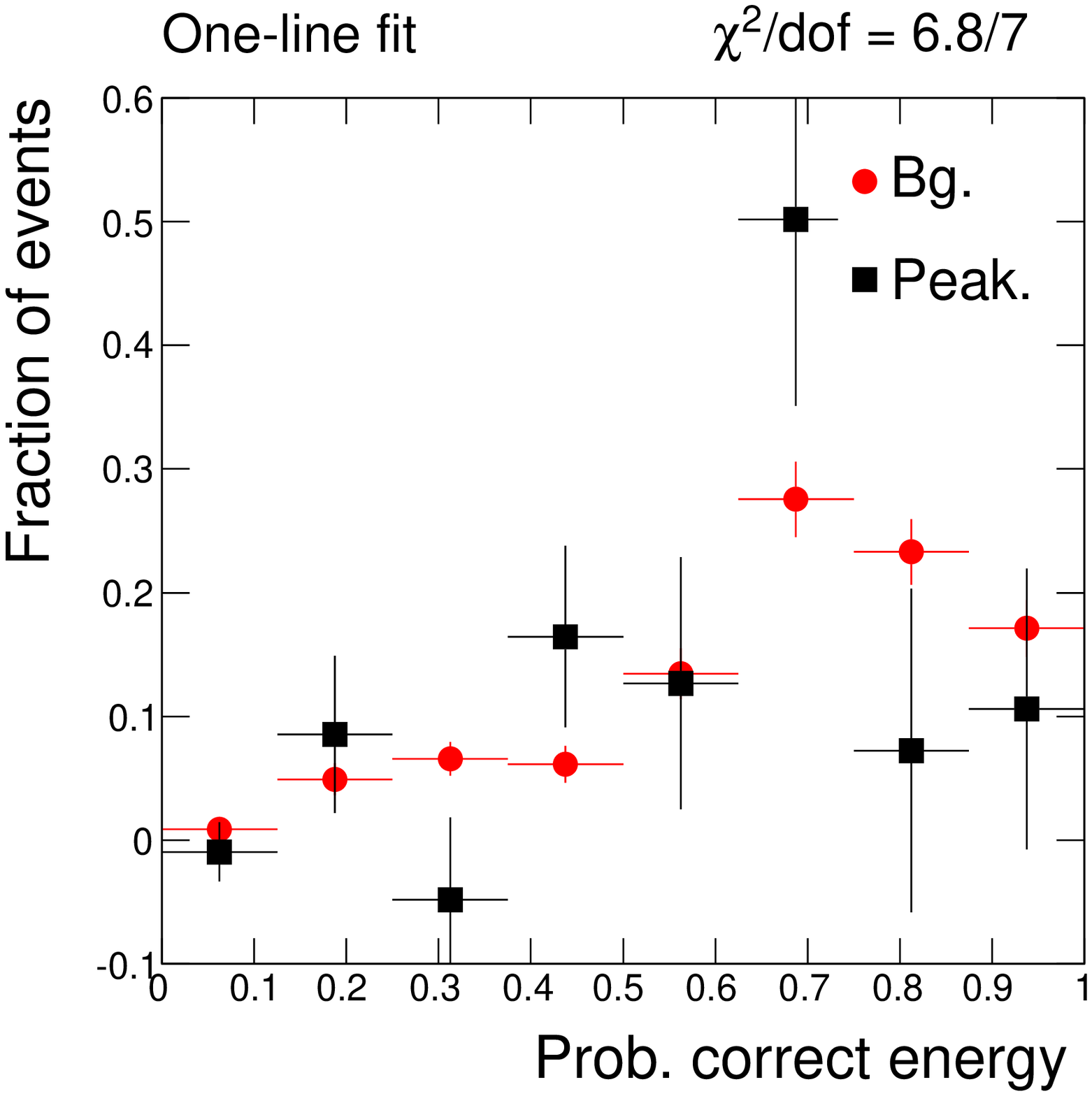}
\includegraphics[width=1.6in]{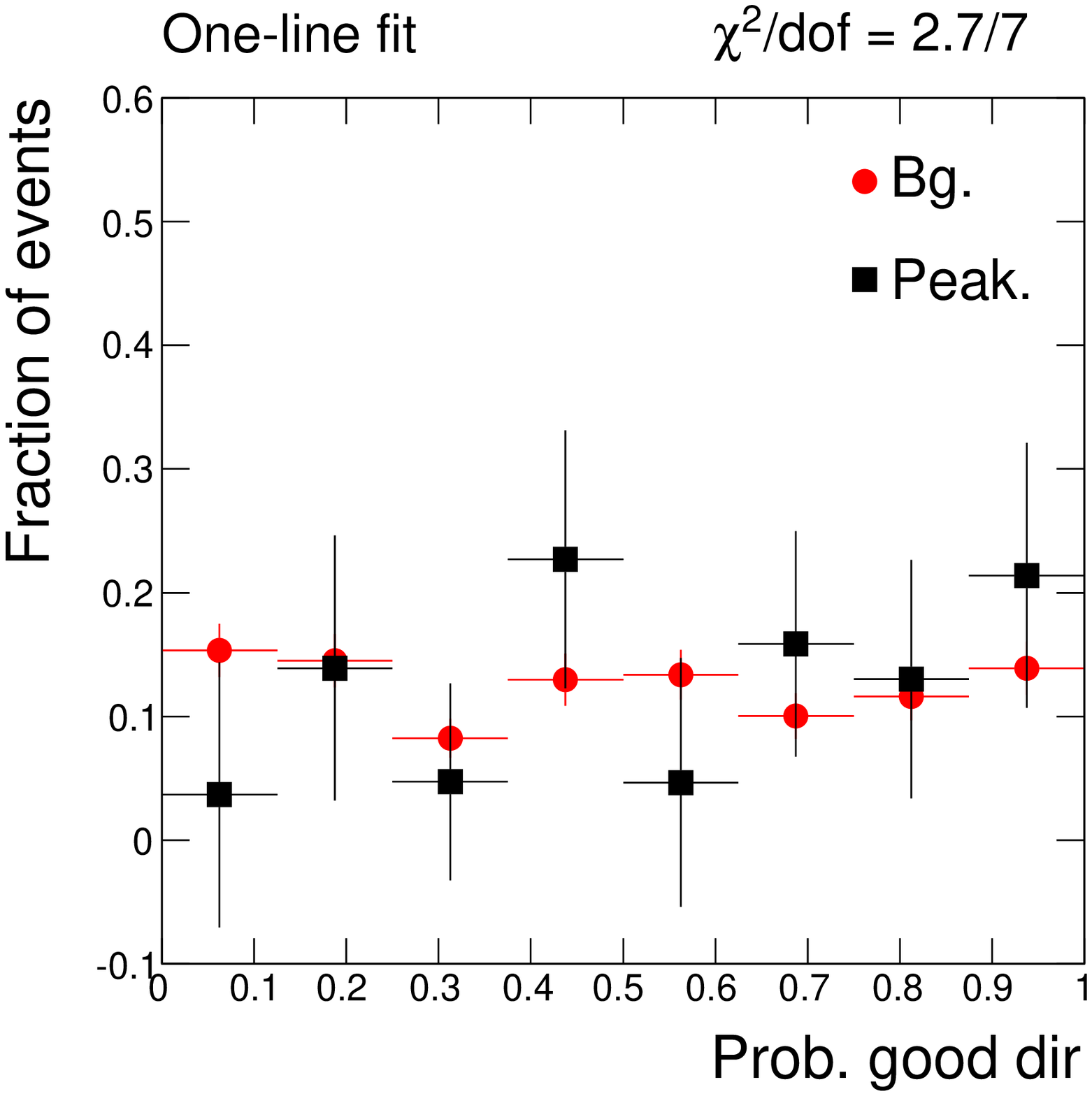}
\caption{Disentangled signal and background distributions.  Left,
  probability of correct energy reconstruction. Right, probability of
  correct angle reconstruction.}
\label{fig:probs1}
\end{figure}

\begin{figure}
\includegraphics[width=1.6in]{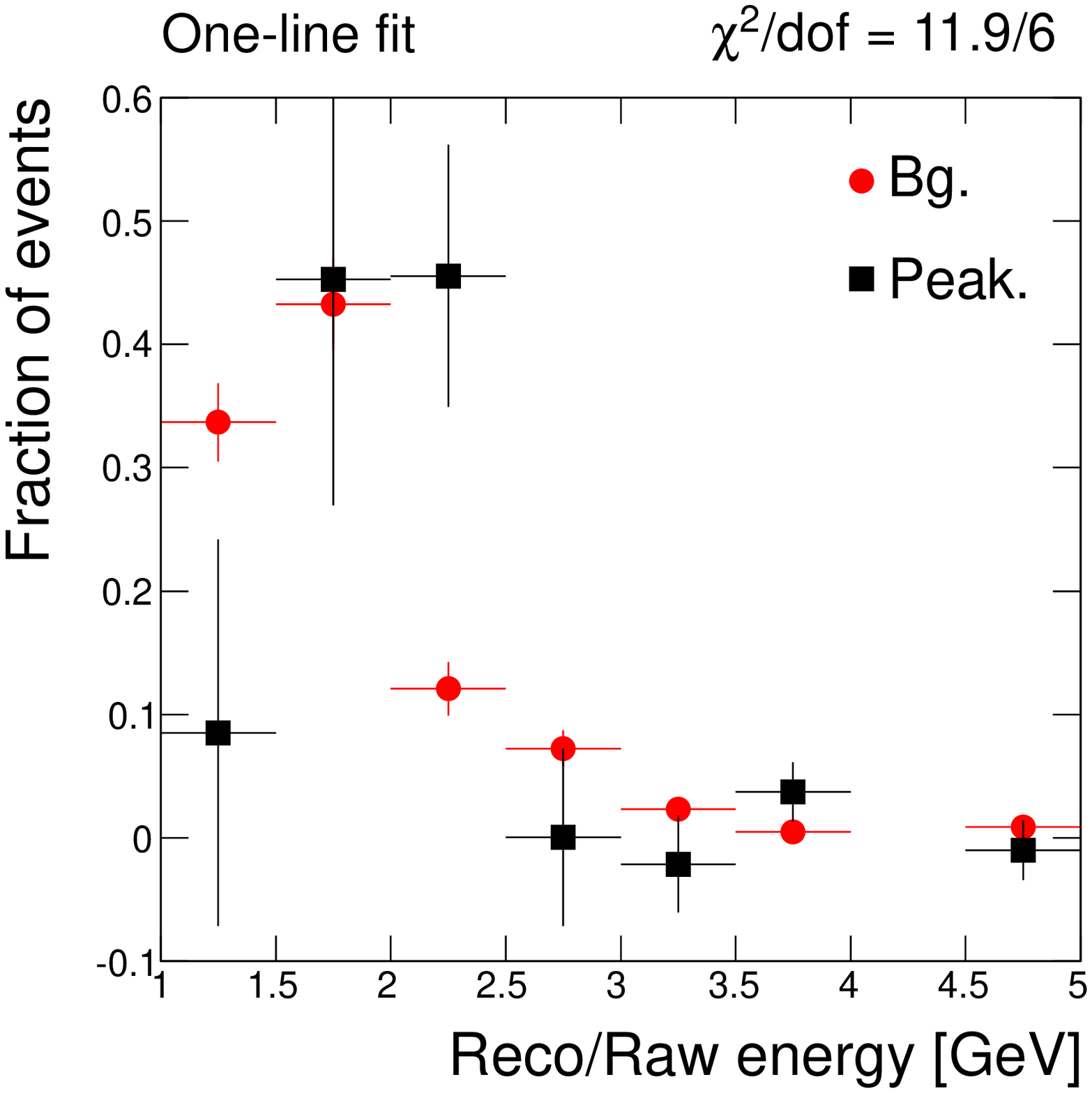}
\includegraphics[width=1.6in]{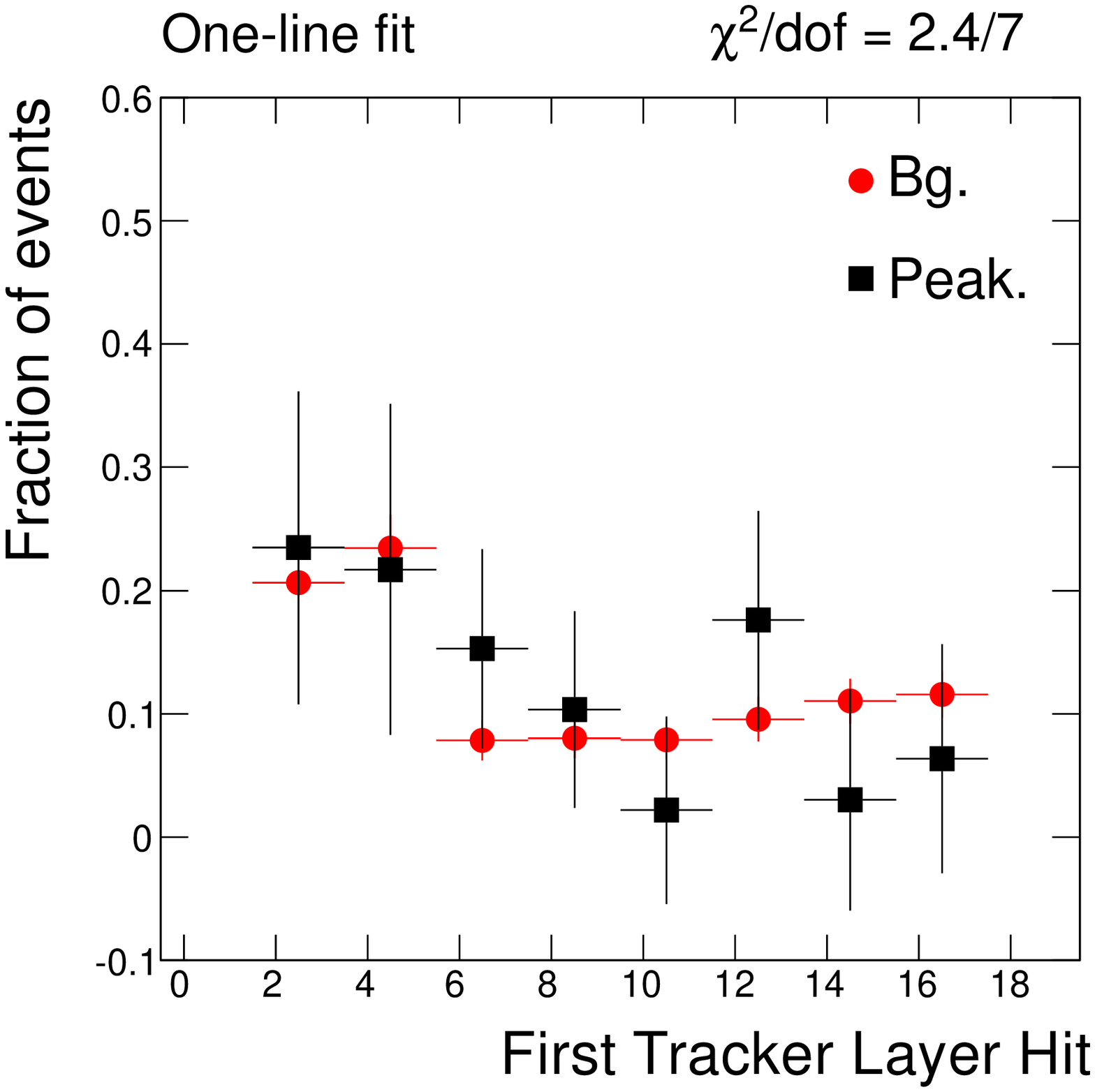}
\caption{Disentangled signal and background distributions.  Left,
  ratio of reconstructed to raw photon energy. Right, the first layer
  of the tracker with a hit.}
\label{fig:ratiotrk1}
\end{figure}

\begin{figure}
\includegraphics[width=1.6in]{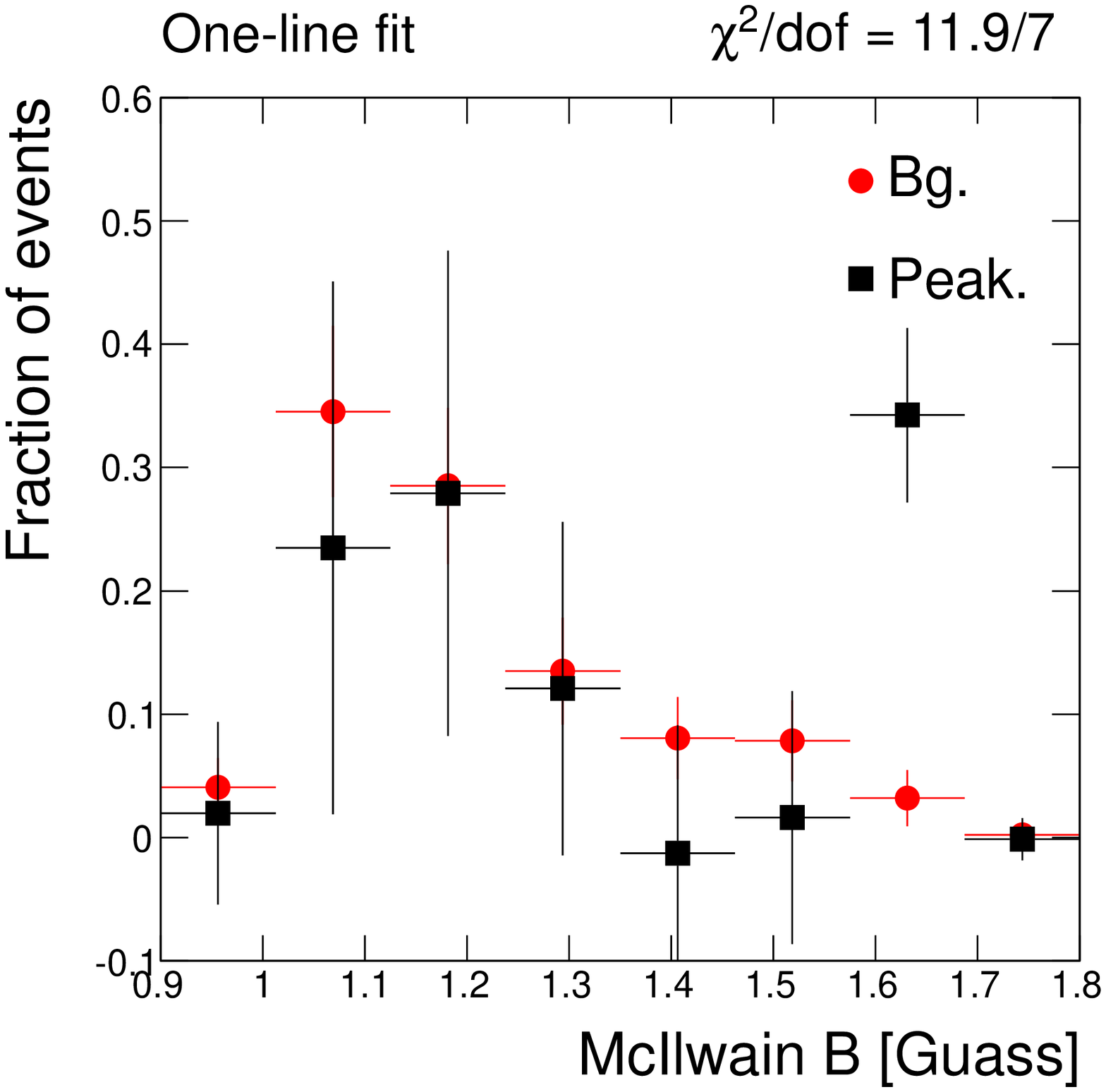}
\includegraphics[width=1.6in]{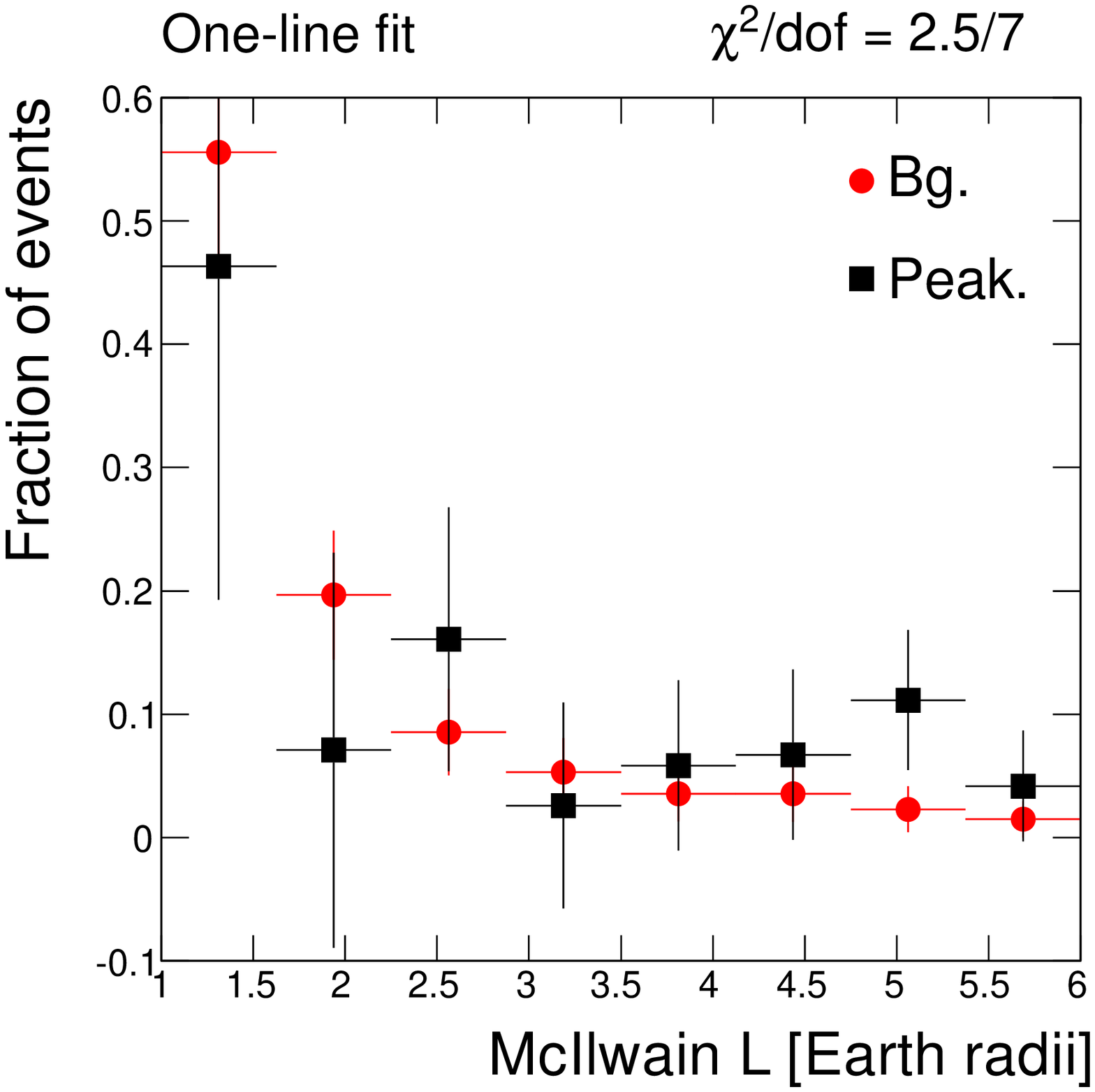}
\caption{Disentangled signal and background distributions.  Left,
  magnetic field strength in terms of the McIlwain $B$
  parameter. Right, the McIlwait $L$ parameter.}
\label{fig:mag1}
\end{figure}

\begin{figure}
\includegraphics[width=1.6in]{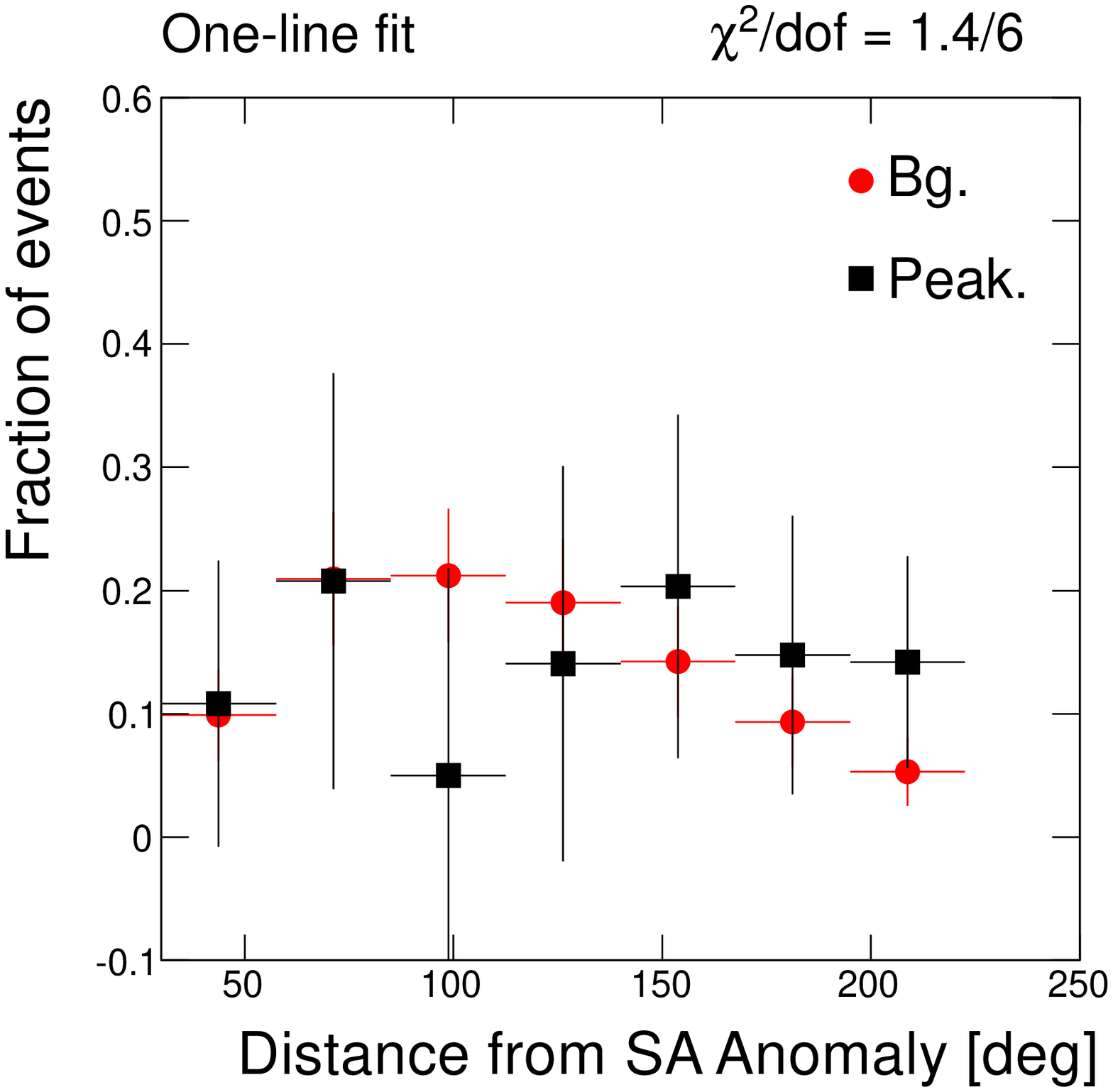}
\includegraphics[width=1.6in]{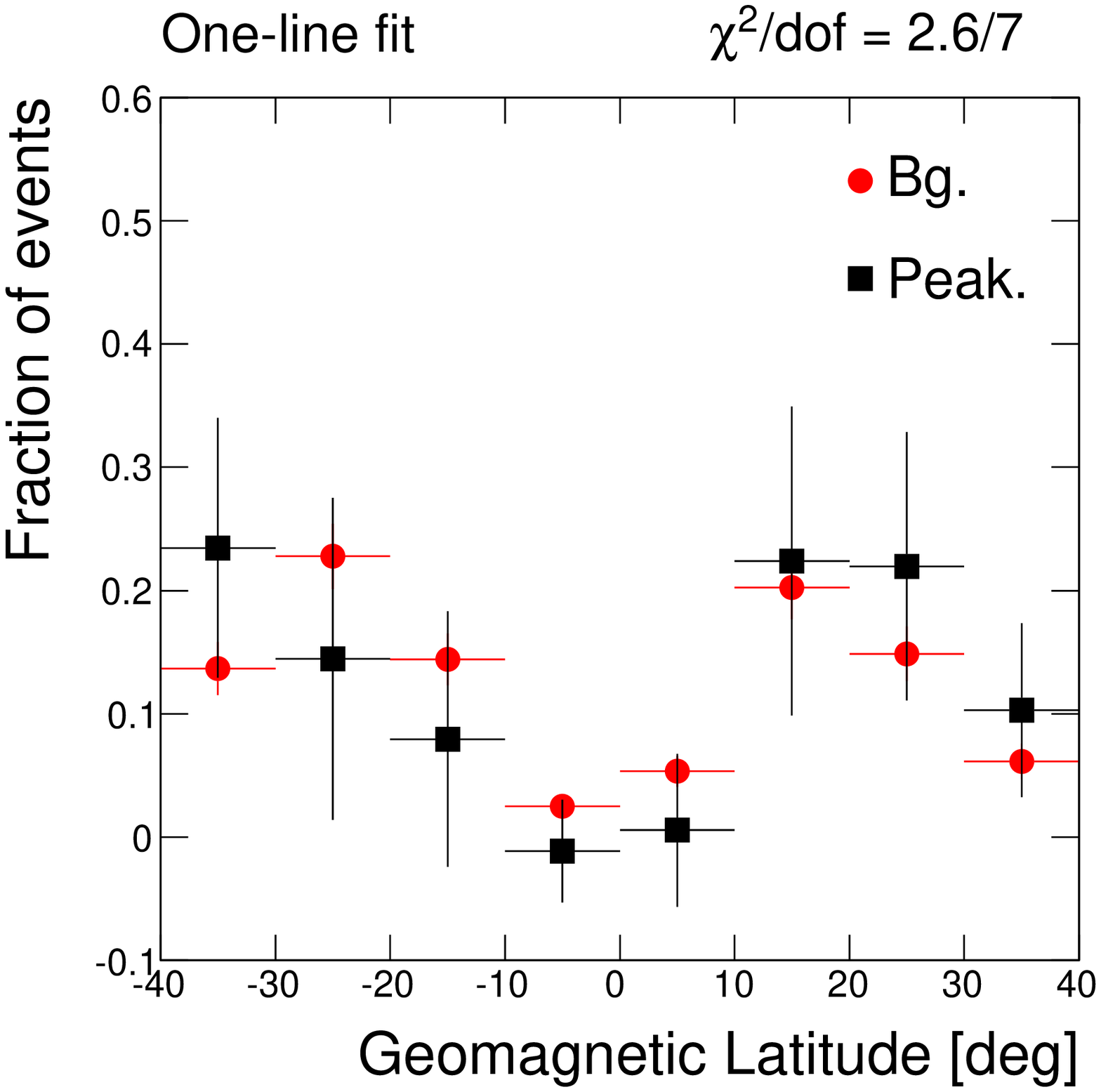}
\caption{Disentangled signal and background distributions. Left,
  the distance in Earth longitude and latitude from the center of the
  South Atlantic Anomaly.  Right, geomagnetic latitude of the
  spacecraft.}
\label{fig:saa1}
\end{figure}

Unfolded distributions of incidence angles are shown in
Fig.~\ref{fig:detang1}. The distributions in galactic coordinates can
be seen in Fig.~\ref{fig:coord1}. Zenith and azimuthal angle distributions are in
Fig.~\ref{fig:zenith1}, and the recorded time and conversion type are
in Fig.~\ref{fig:timeback1}. Energy and direction reconstruction
quality are in Fig.~\ref{fig:probs1} and the reconstructed/raw energy
ratio as well as the first layer of the tracker with a hit are shown
in Fig.~\ref{fig:ratiotrk1}. The magnetic field parameters are shown
in Fig.~\ref{fig:mag1} and the distance from the South Atlantic anomoly
and the geomagnetic latitude are shown in Fig.~\ref{fig:saa1}.

 In each case, we compare the
distributions quantitatively by calculating the $\chi^2/$dof between
the peak and background distributions, shown in Table~\ref{tab:stat}. As the signal and background
weights are anti-correlated, this is calculated as

\[ \chi^2 =
\sum_{\textrm{bin}\ i}\frac{(N^i_{\textrm{peak}}-N^i_{\textrm{bg.}})^2}{(\Delta N^i_{\textrm{peak}}+\Delta
  N^i_{\textrm{bg.}})^2}\]

\noindent
where $N^i_{\textrm{peak}}$ is the sum of the weights
$\sum sP_{\textrm{peak}}$ in that bin, and $\Delta
N^i_{\textrm{peak}}$ is calculated from toy simulations which estimate
the expected variance of the measurement in each bin.  Similar expressions apply for the
background uncertainties.

\subsection{Double-Line Analysis}

It has been suggested~\cite{twolines} that there may be two photon
lines, the $\gamma\gamma$ feature being accompanied by a feature due
to $\gamma Z$ production, which would be at lower $E_\gamma$ (see Eq (1)).

\begin{figure}
\includegraphics[width=3in]{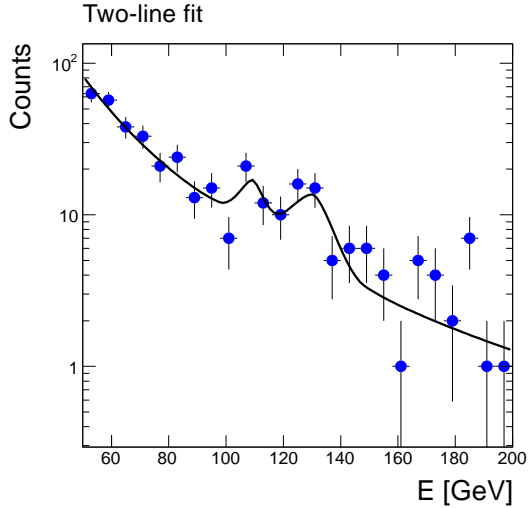}
\caption{Energy of Fermi-LAT photons with signal plus background fit  in the
  double-line analysis at $E_\gamma=110$ GeV and  $E_\gamma=130$ GeV.}
\label{fig:fit2}
\end{figure}

We modify the signal pdf to include two lines, one at 110 GeV and one
at 130 GeV (the results are not sensitive to the precise position of
the second line). We allow the two line features to float independently, but
in the {\sc sPlots} analysis we treat them together as a single pdf once
their relative normalization has been fixed by the fit. The result of
the fit can be seen in Figure~\ref{fig:fit2}.   Note, however, that
systematic or instrumental issues which cause  features in the energy
spectrum at 110 GeV and 130 GeV may not be manifested in the same regions of the instrumental
variables, and so may not add coherently.

\begin{figure}
\includegraphics[width=1.6in]{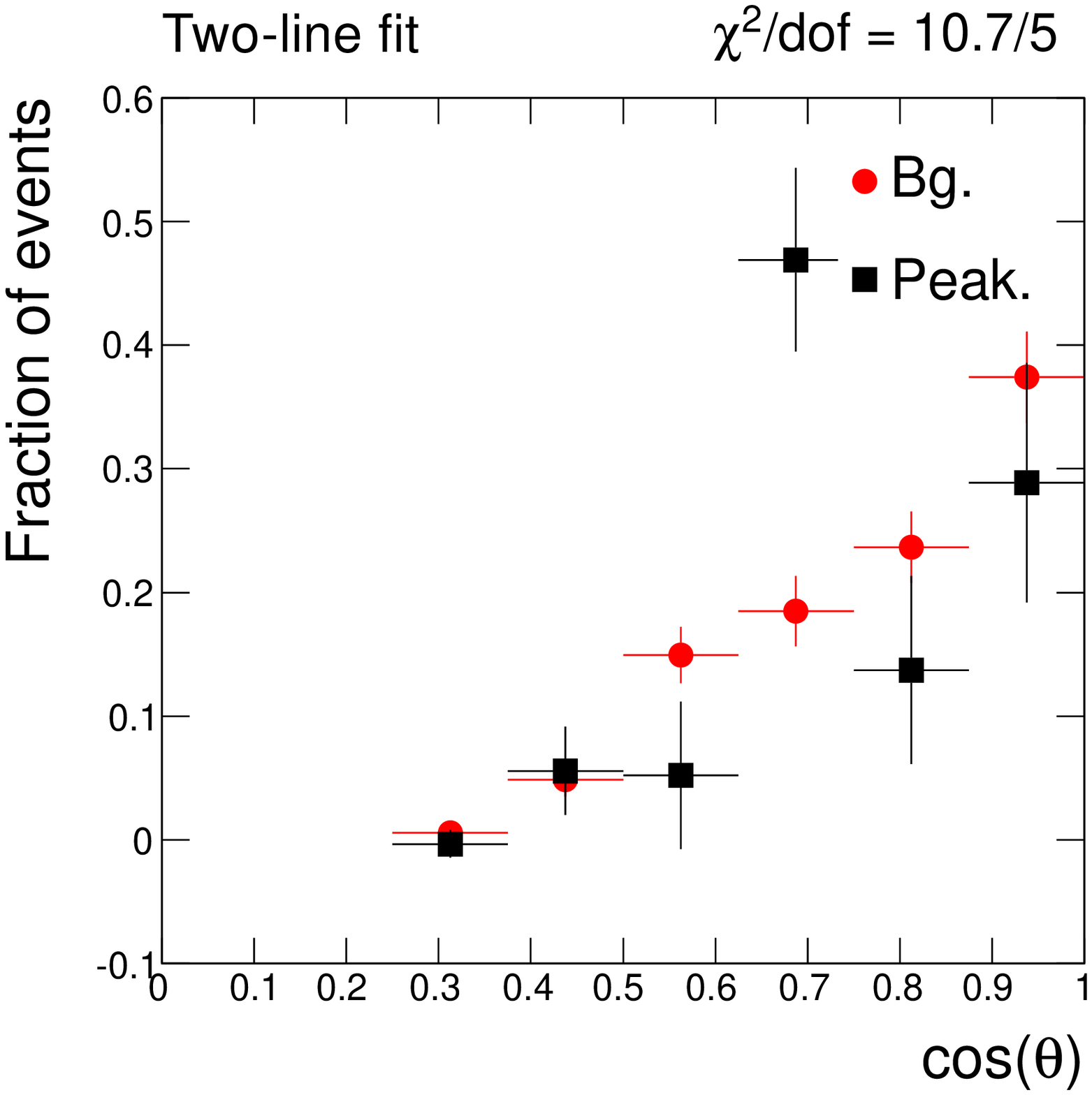}
\includegraphics[width=1.6in]{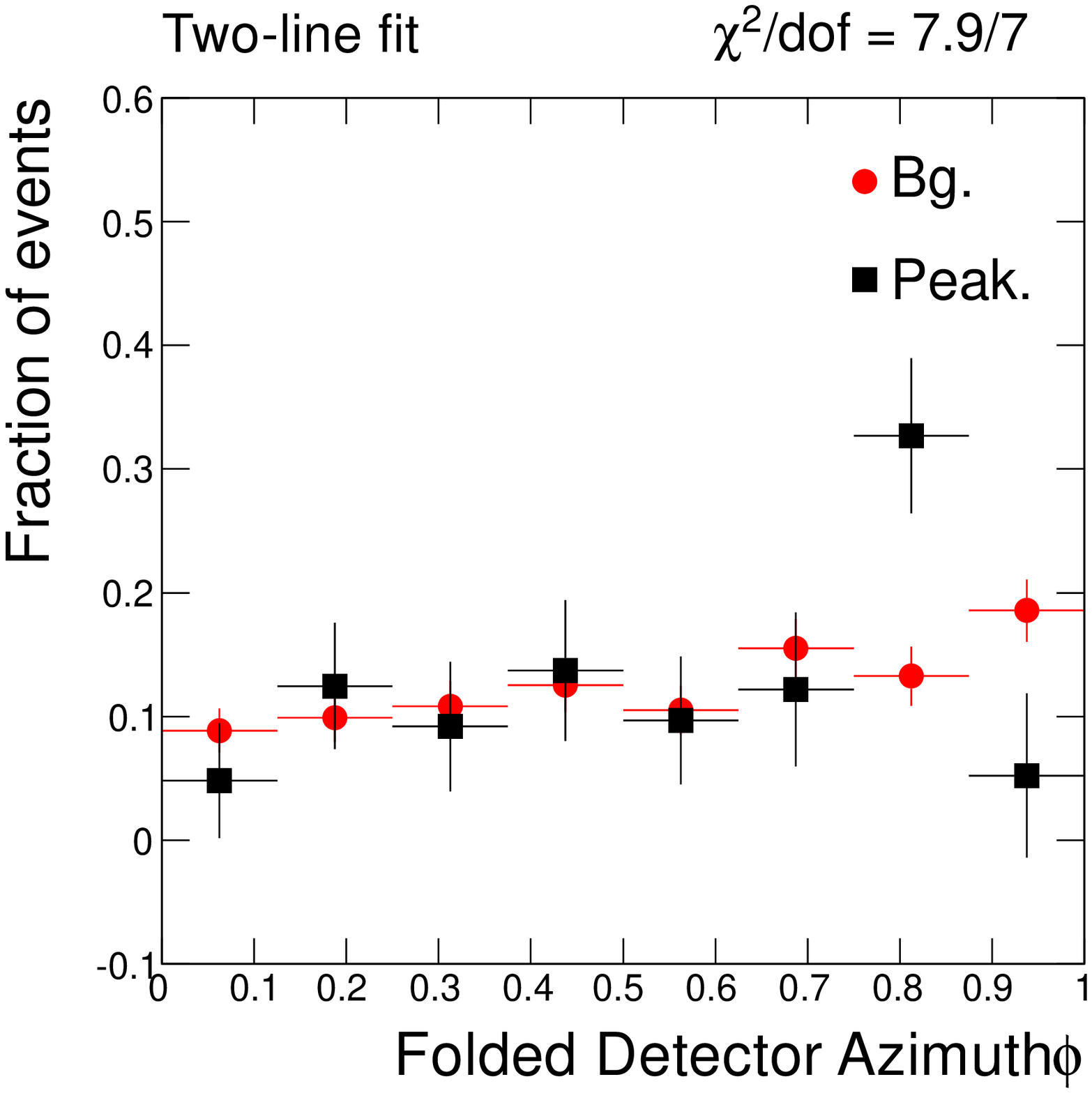}
\caption{Disentangled signal and background distributions  in the
  double-line analysis. Left,  $\cos(\theta)$ where
  $\theta$ is the photon incidence angle relative  to a line normal
  the Fermi-LAT face. Right, $\phi$, the photon incidence angle
  relative to the sun-facing side~\cite{fermidefs}.}
\label{fig:detang2}
\end{figure}

\begin{figure}
\includegraphics[width=1.6in]{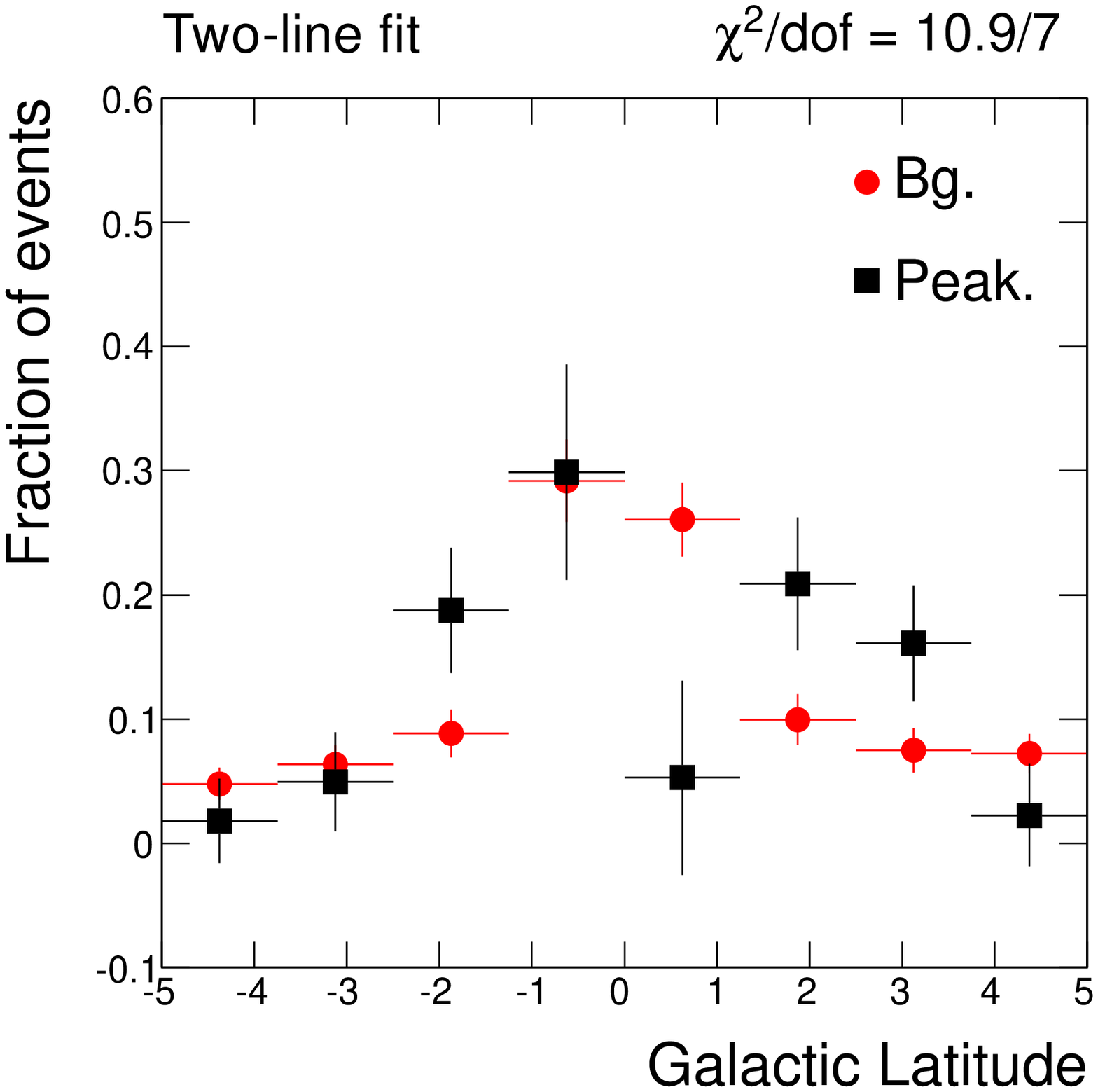}
\includegraphics[width=1.6in]{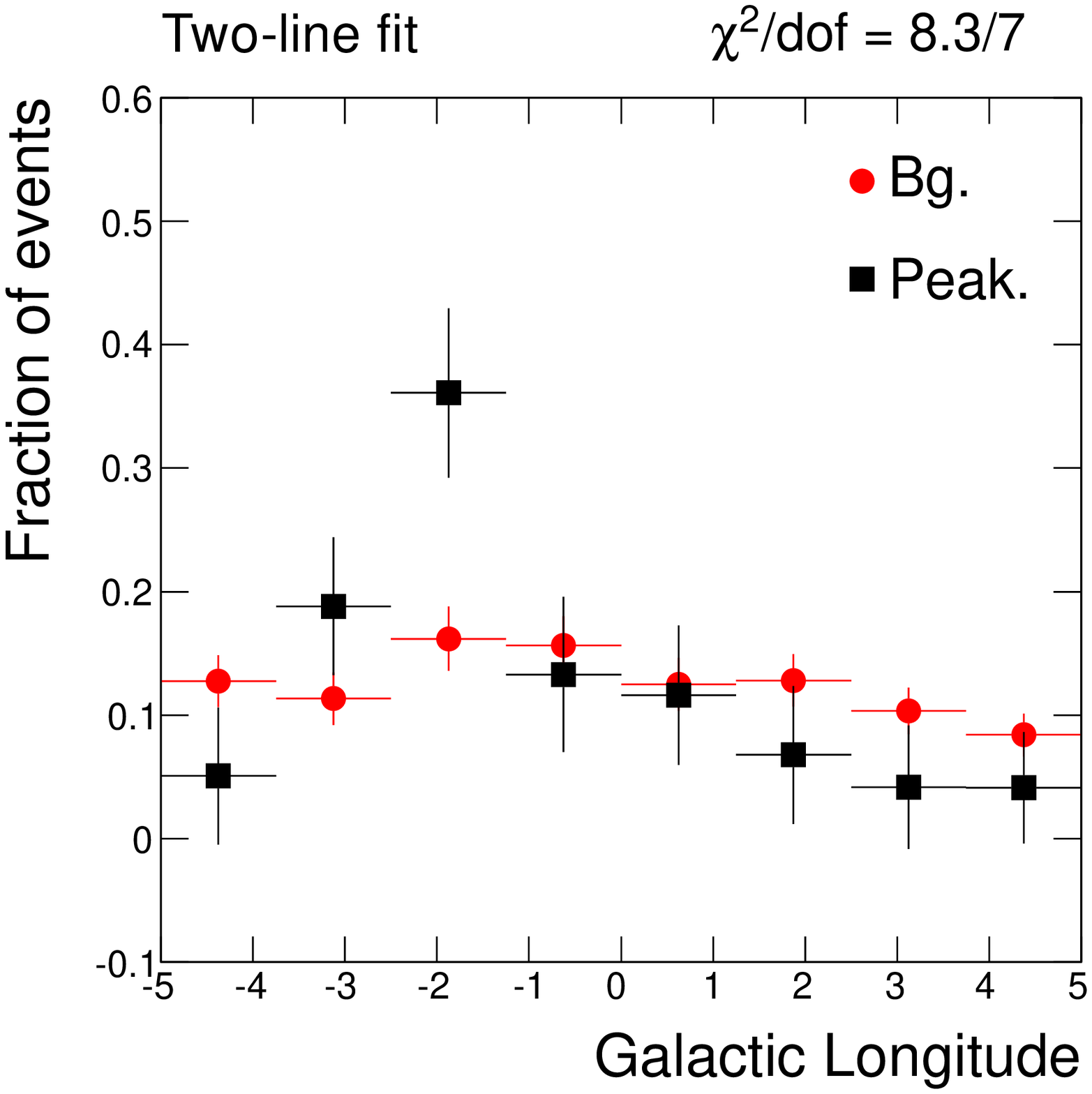}
\caption{Disentangled signal and background distributions  in the
  double-line analysis in galactic
  coordinates, $b$ (latitude) and $l$
  (longitude)~\cite{fermidefs}. No smoothing has been applied.}
\label{fig:coord2}
\end{figure}

\begin{figure}
\includegraphics[width=1.6in]{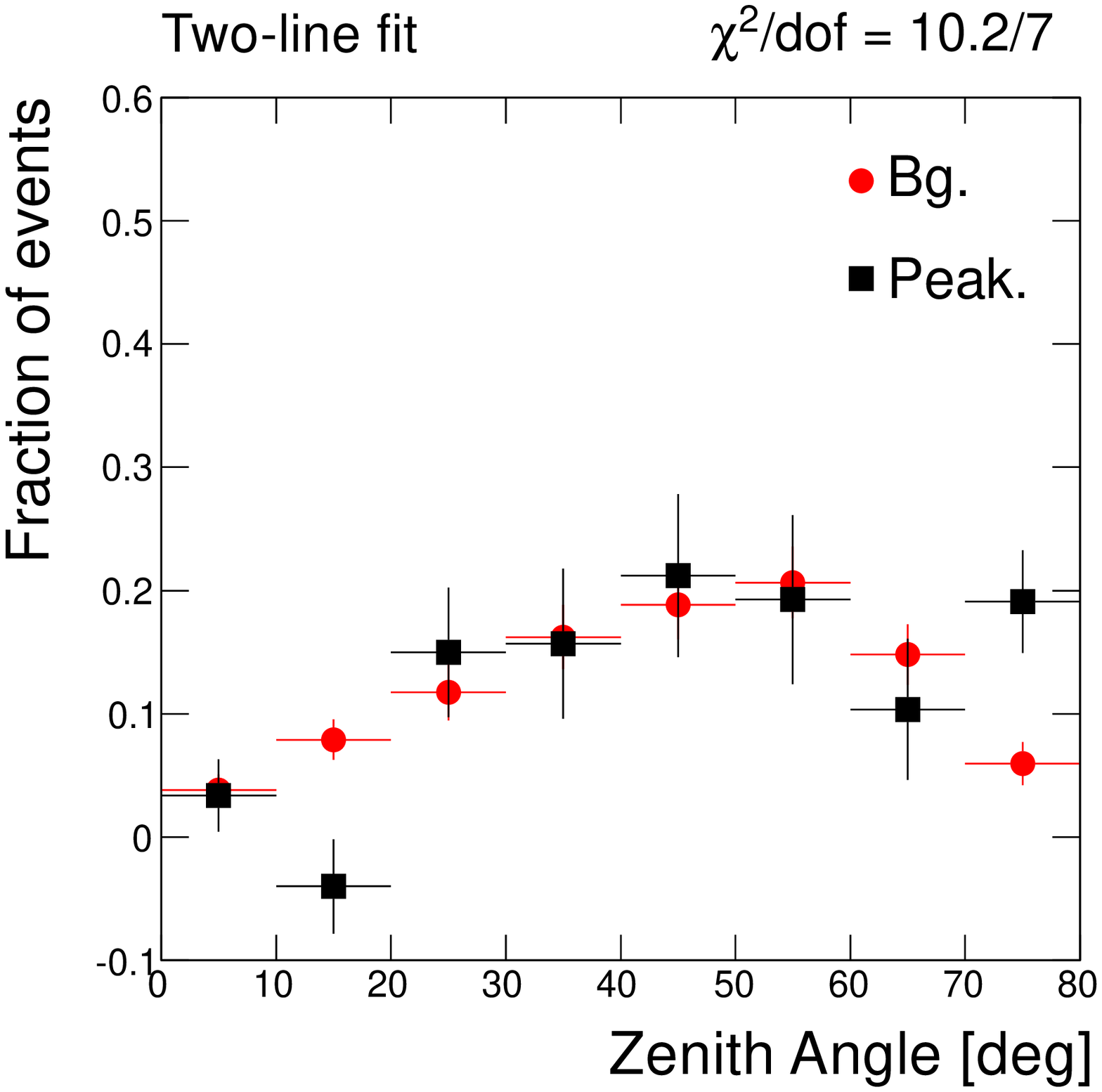}
\includegraphics[width=1.6in]{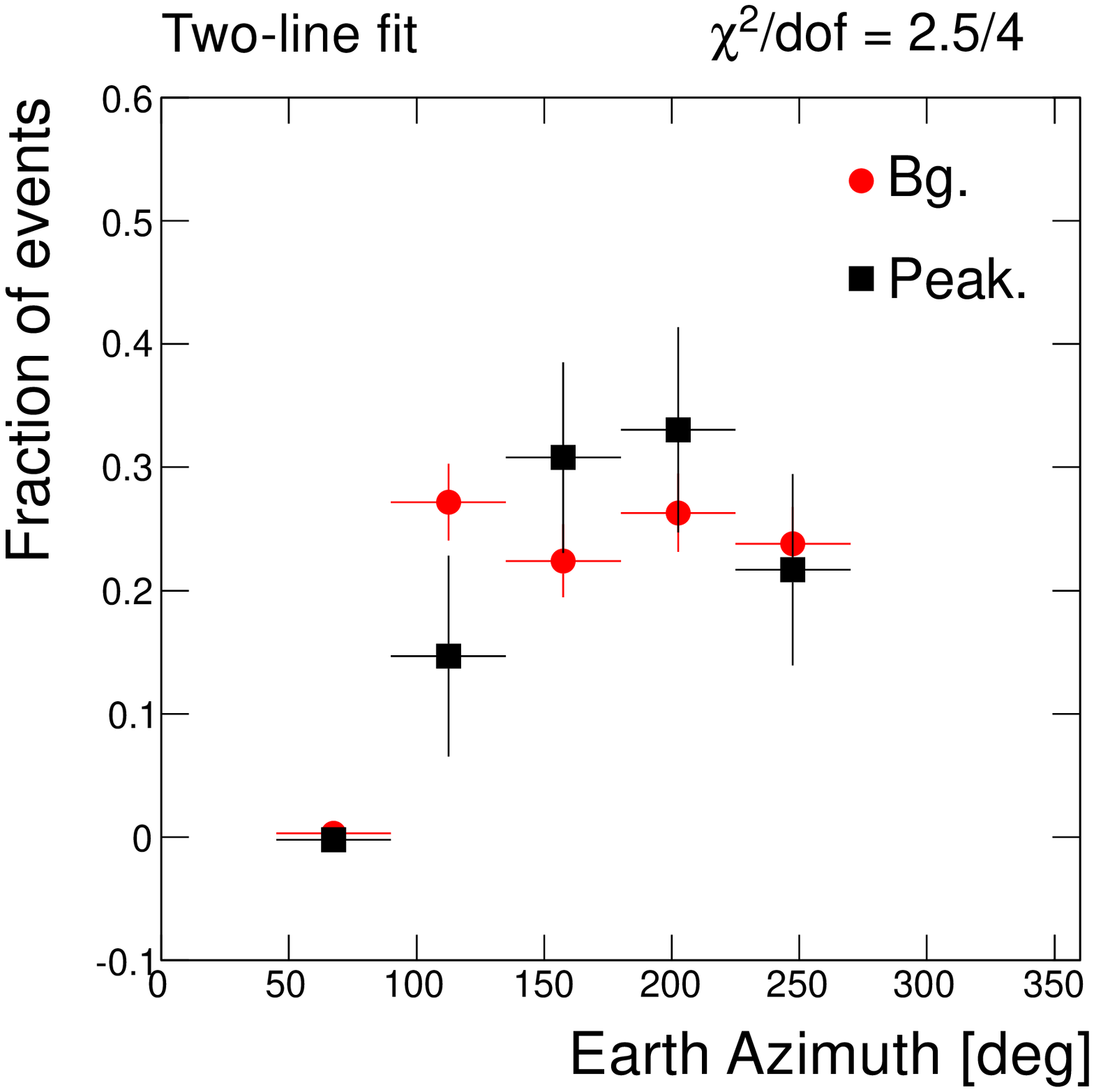}
\caption{Disentangled signal and background distributions in the
  double-line analysis. Left, angle
  between the reconstructed photon direction and the zenith line,
  which passed through the earth and Fermi's center of mass. Right,
  the earth azimuth angle. }
\label{fig:zenith2}
\end{figure}

\begin{figure}
\includegraphics[width=1.6in]{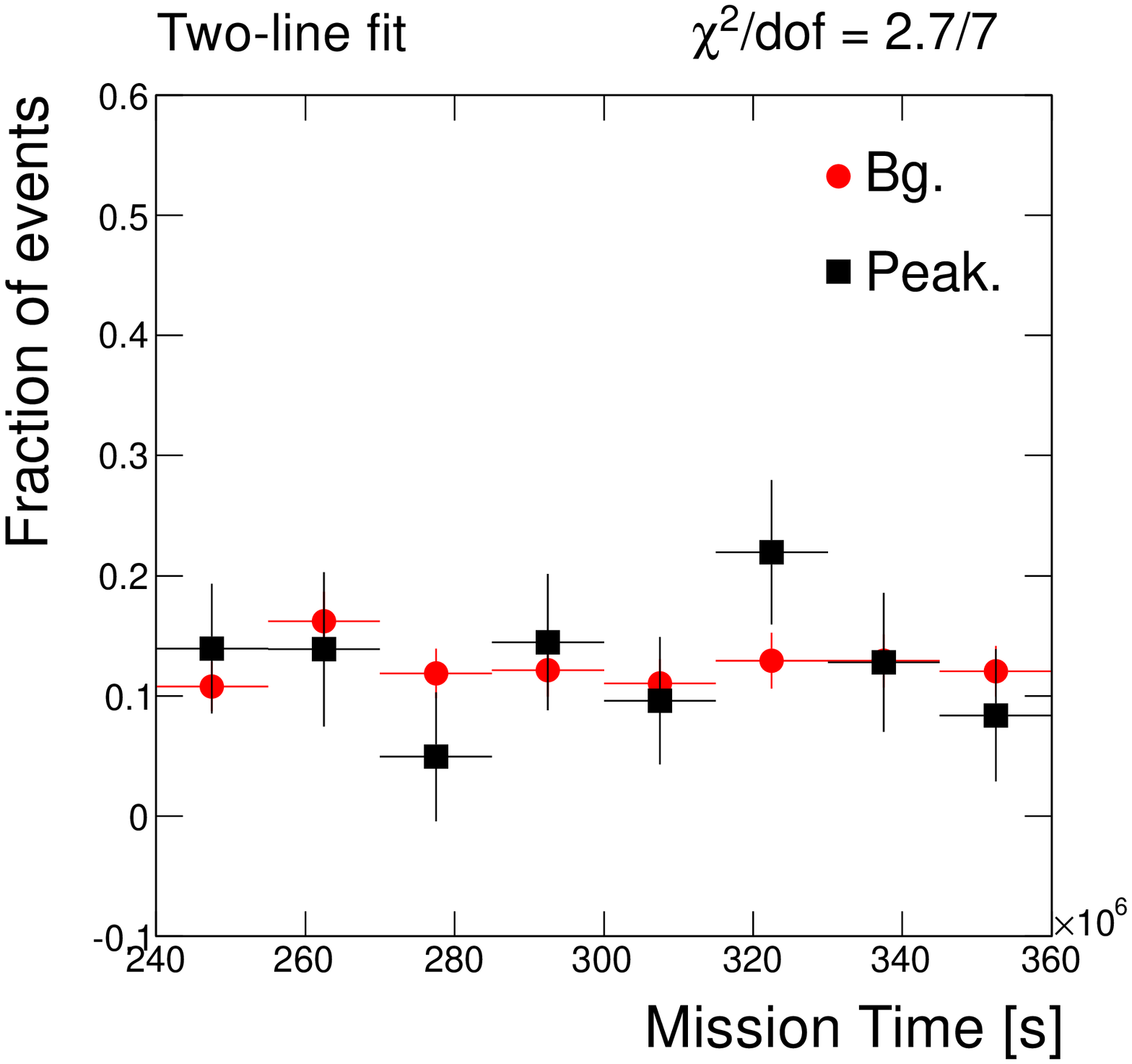}
\includegraphics[width=1.6in]{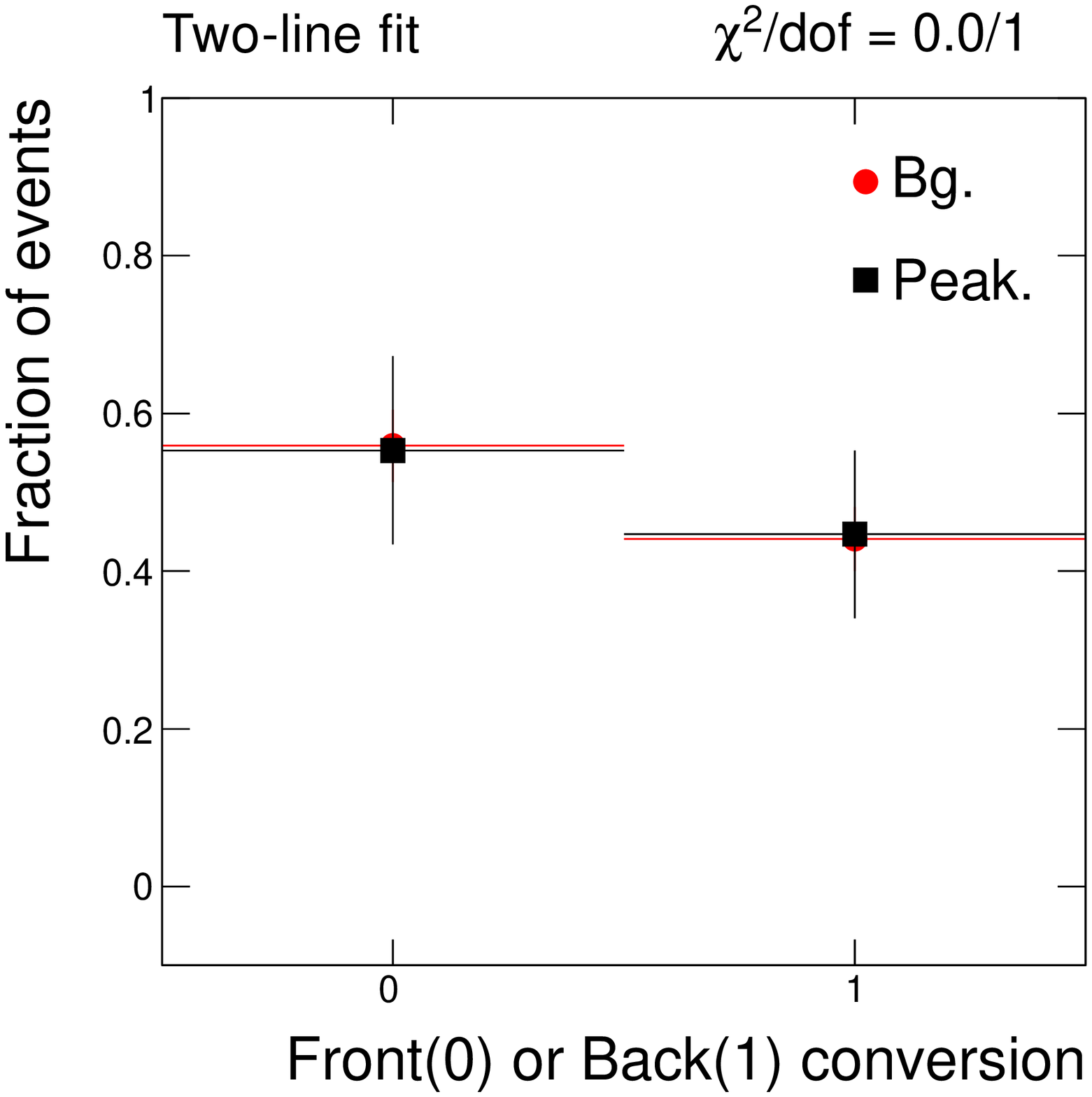}
\caption{Disentangled signal and background distributions in the
  double-line analysis.  Left, the mission elapsed time since Jan
  1 2001~\cite{fermidefs}. Right, 
  fraction of events in which the pair production is induced in the
  front (thin) or back (thick) layers of the tracker.}
\label{fig:timeback2}
\end{figure}

\begin{figure}
\includegraphics[width=1.6in]{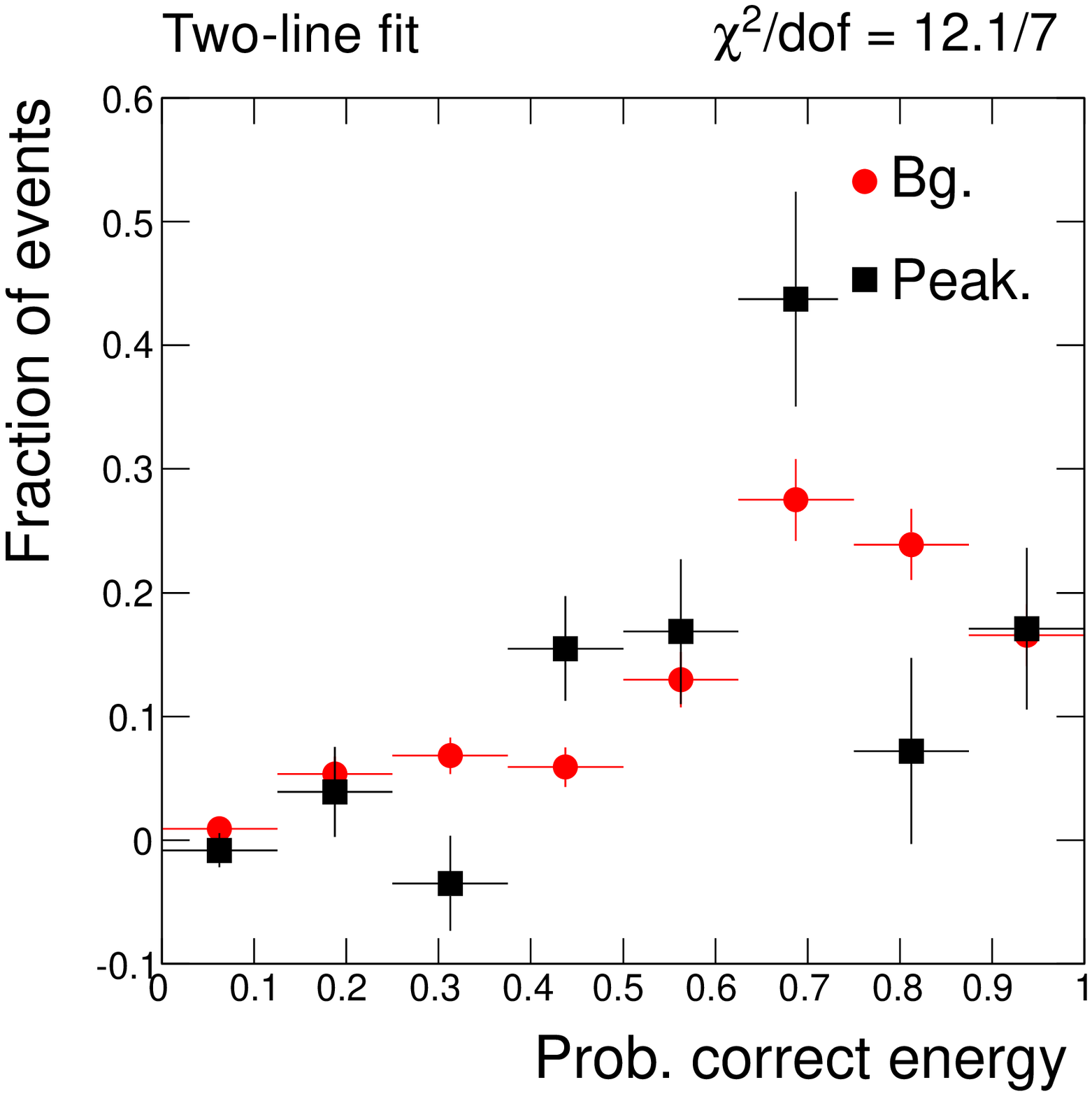}
\includegraphics[width=1.6in]{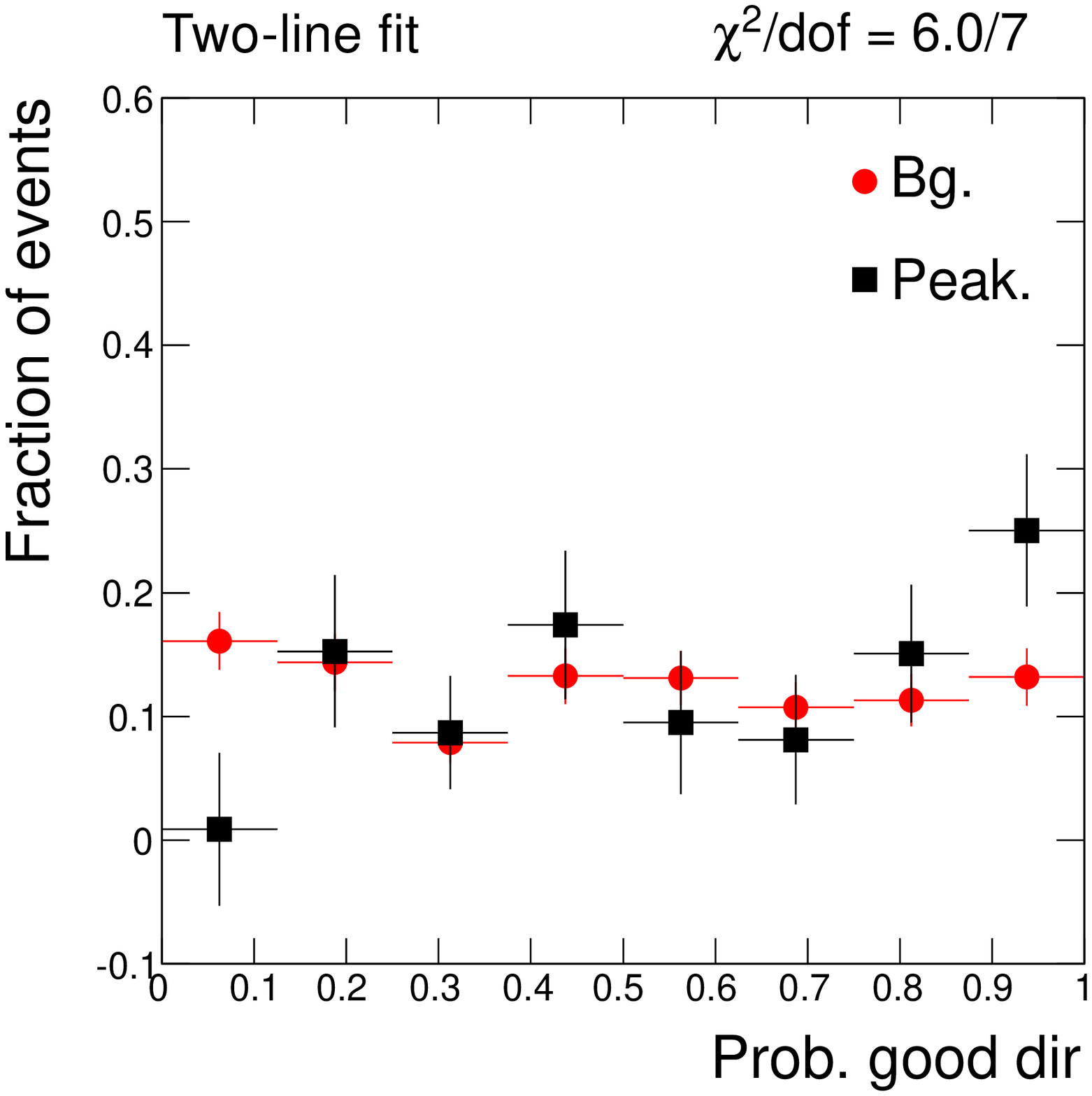}
\caption{Disentangled signal and background distributions.  Left,
  probability of correct energy reconstruction. Right, probability of
  correct angle reconstruction.}
\label{fig:probs2}
\end{figure}

\begin{figure}
\includegraphics[width=1.6in]{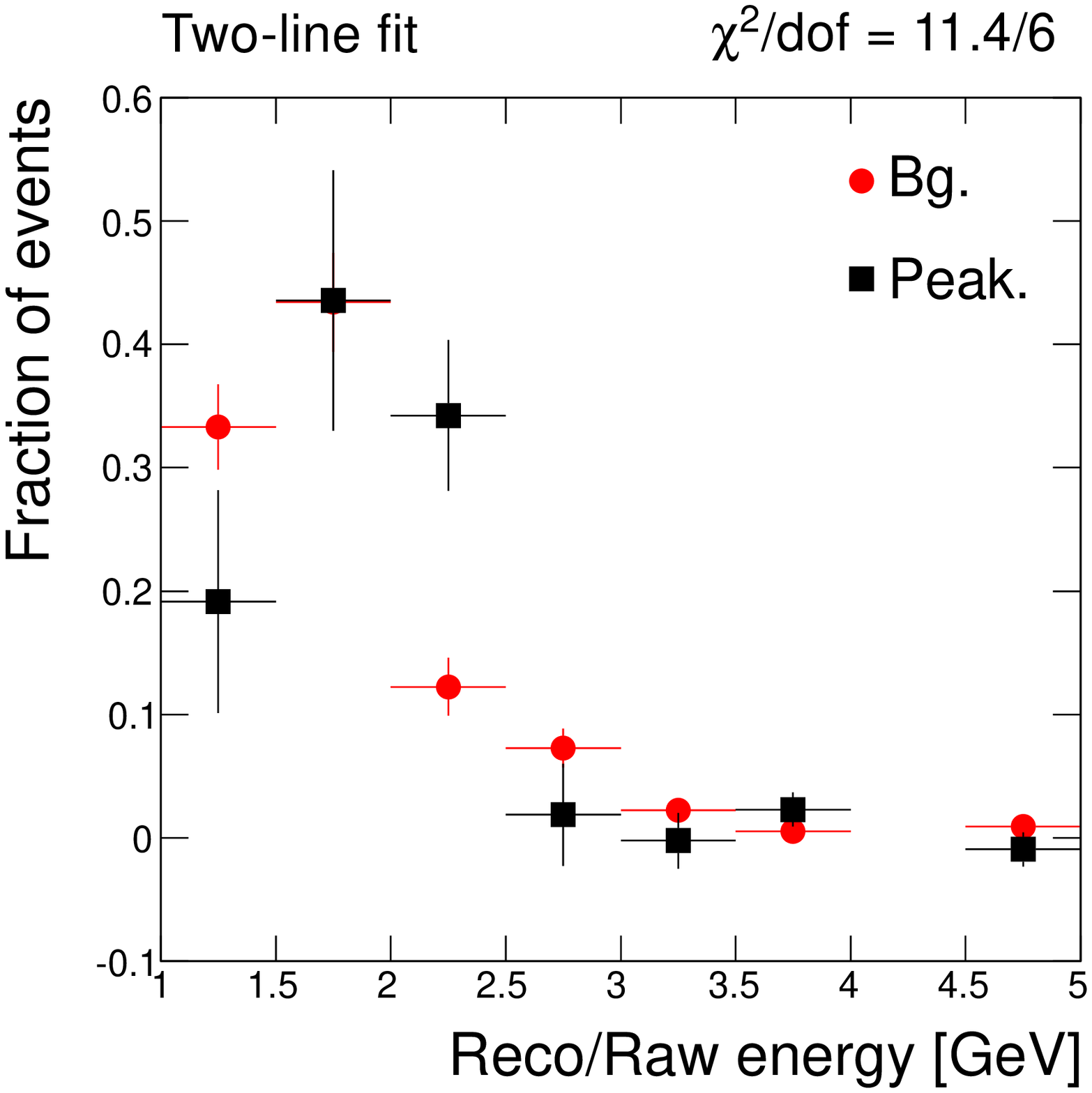}
\includegraphics[width=1.6in]{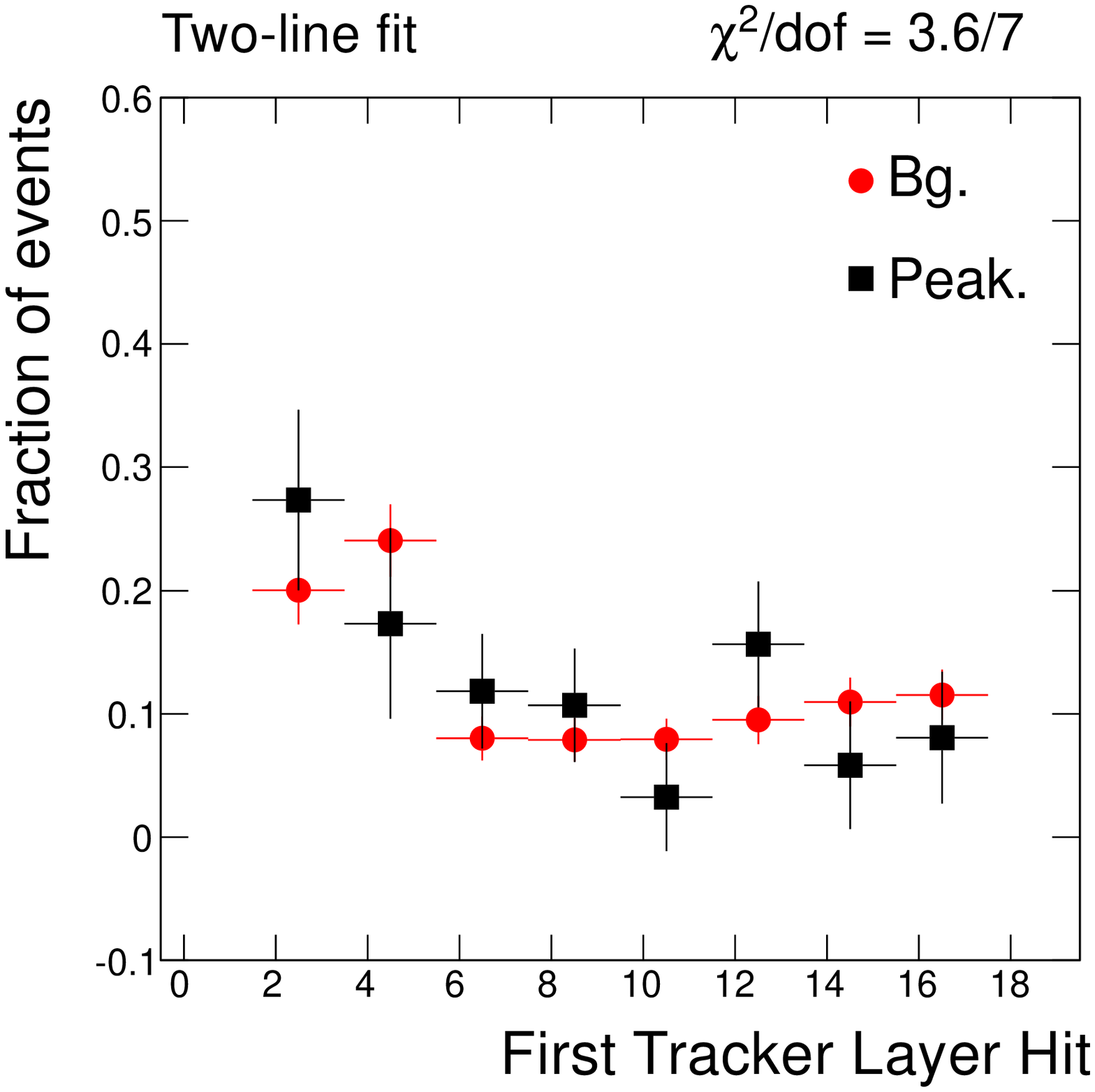}
\caption{Disentangled signal and background distributions.  Left,
  ratio of reconstructed to raw photon energy. Right, the first layer
  of the tracker with a hit.}
\label{fig:ratiotrk2}
\end{figure}

\begin{figure}
\includegraphics[width=1.6in]{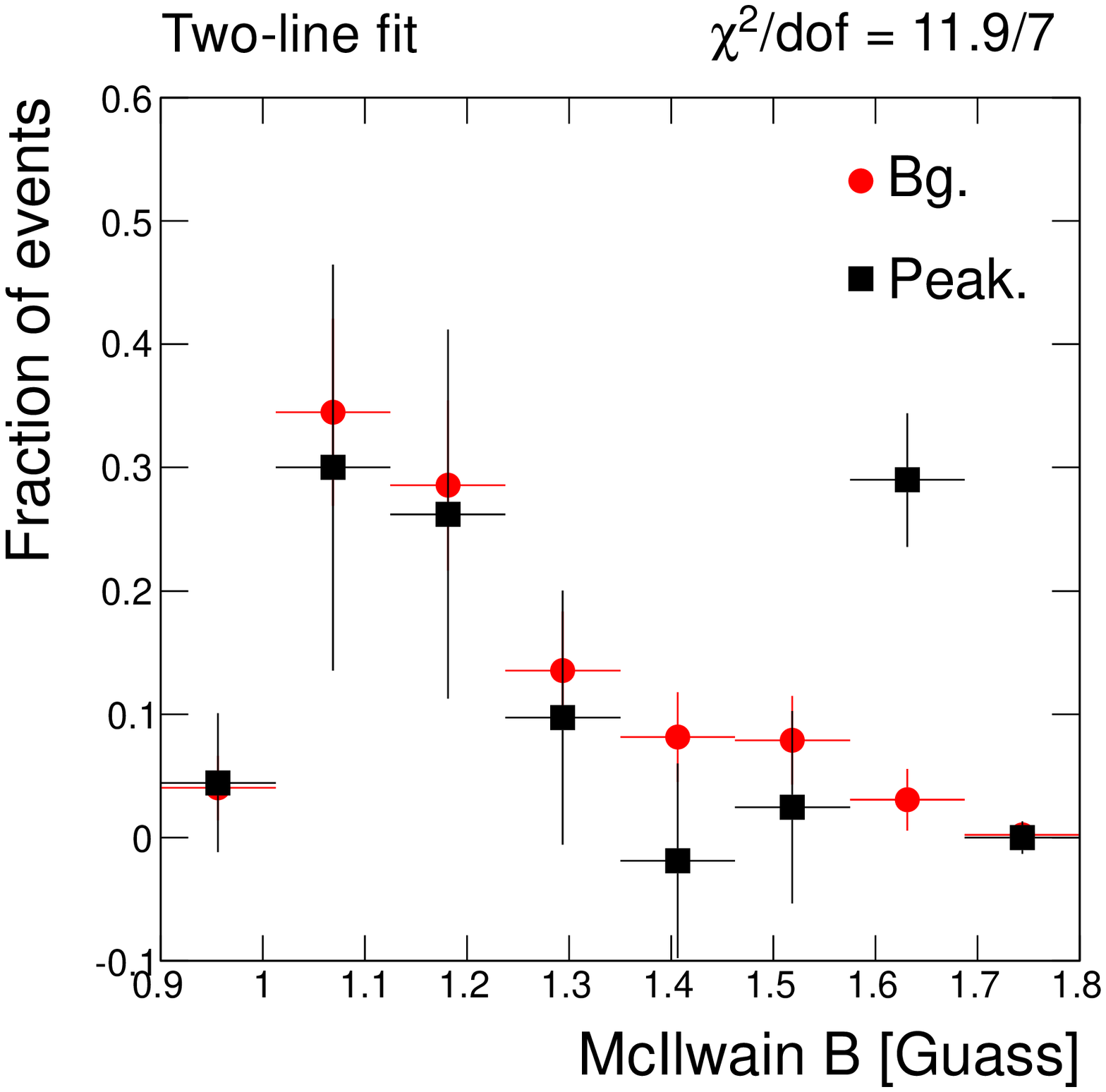}
\includegraphics[width=1.6in]{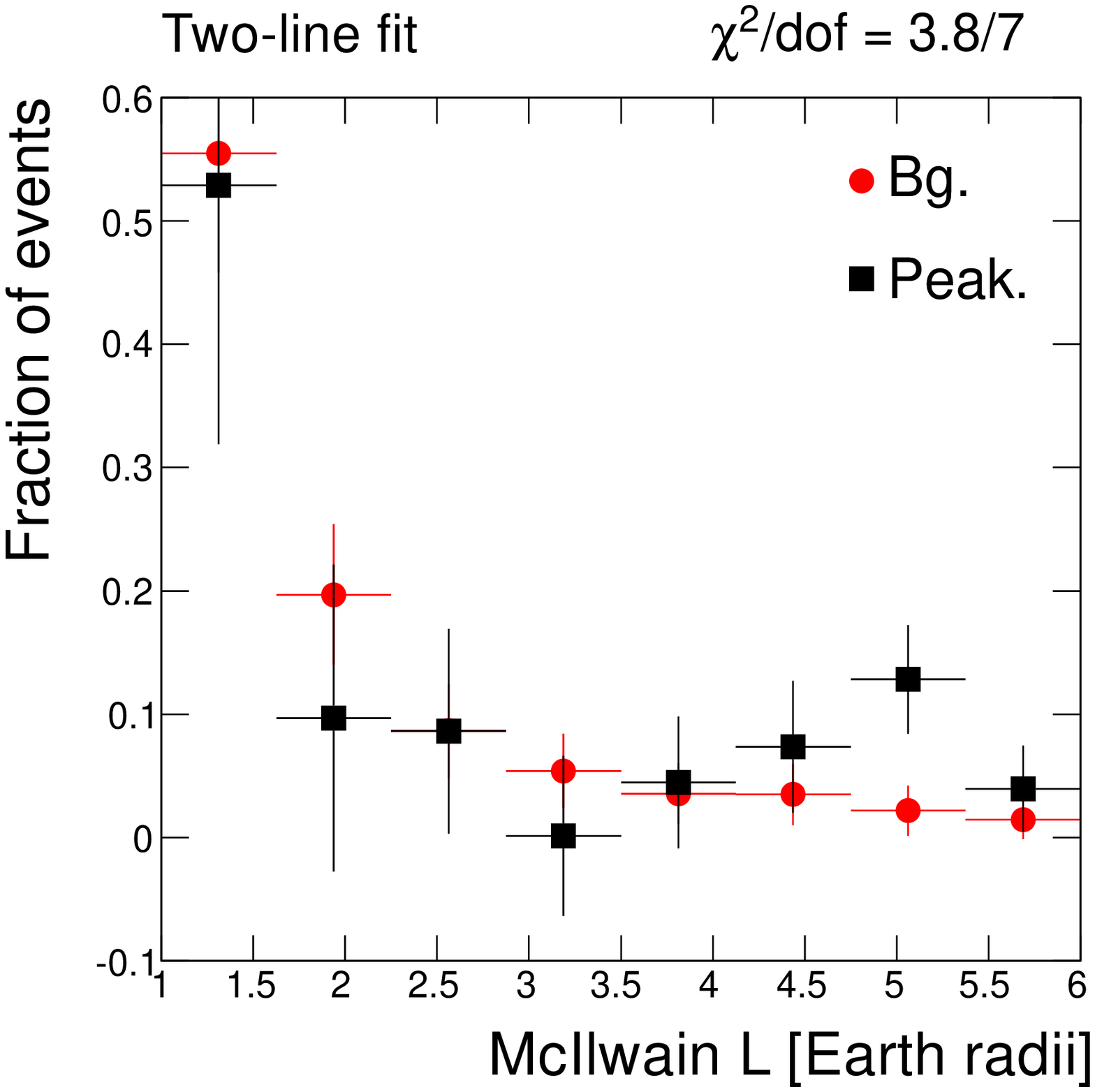}
\caption{Disentangled signal and background distributions.  Left,
  magnetic field strength in terms of the McIlwain $B$
  parameter. Right, the McIlwait $L$ parameter.}
\label{fig:mag2}
\end{figure}

\begin{figure}
\includegraphics[width=1.6in]{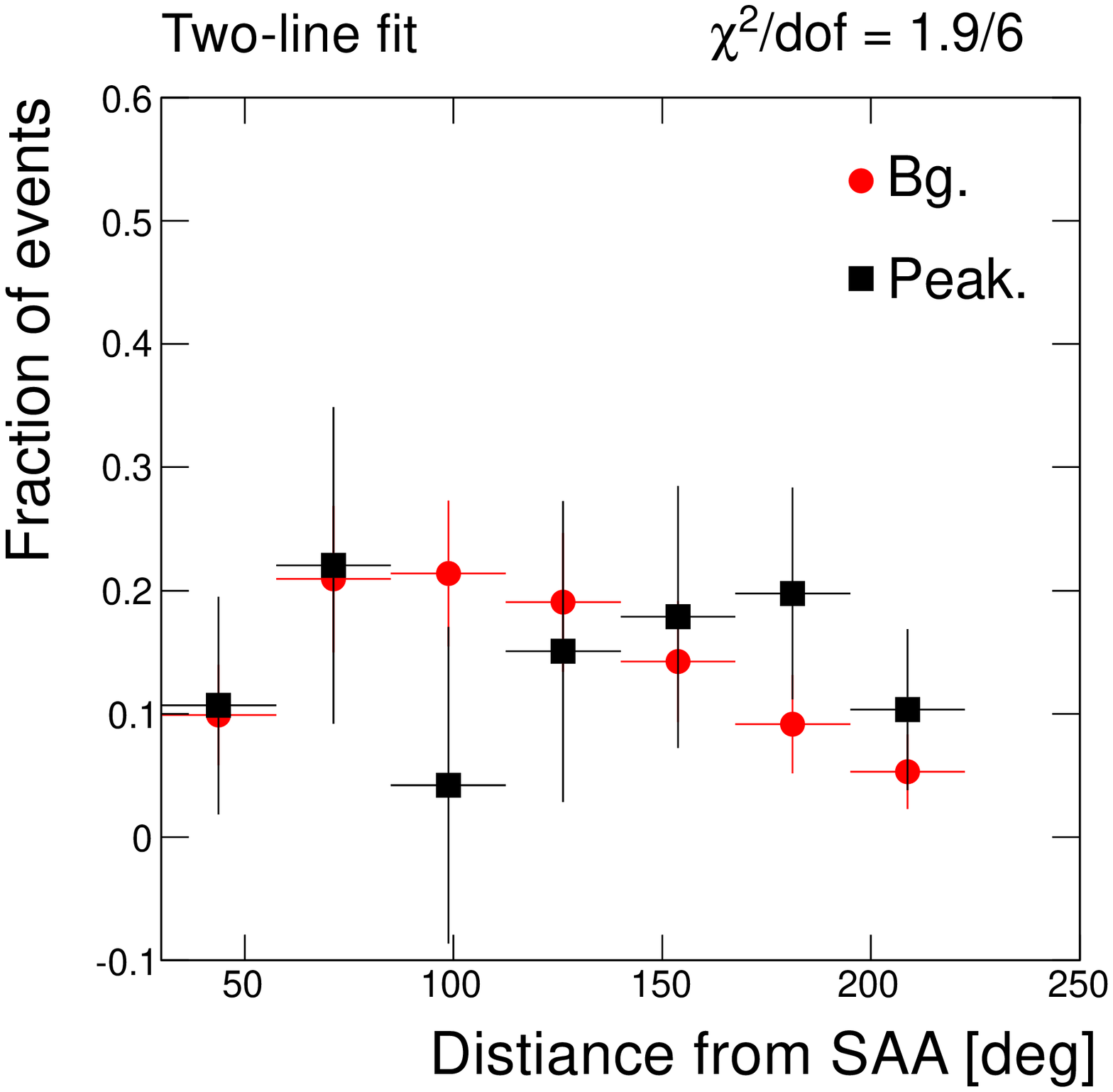}
\includegraphics[width=1.6in]{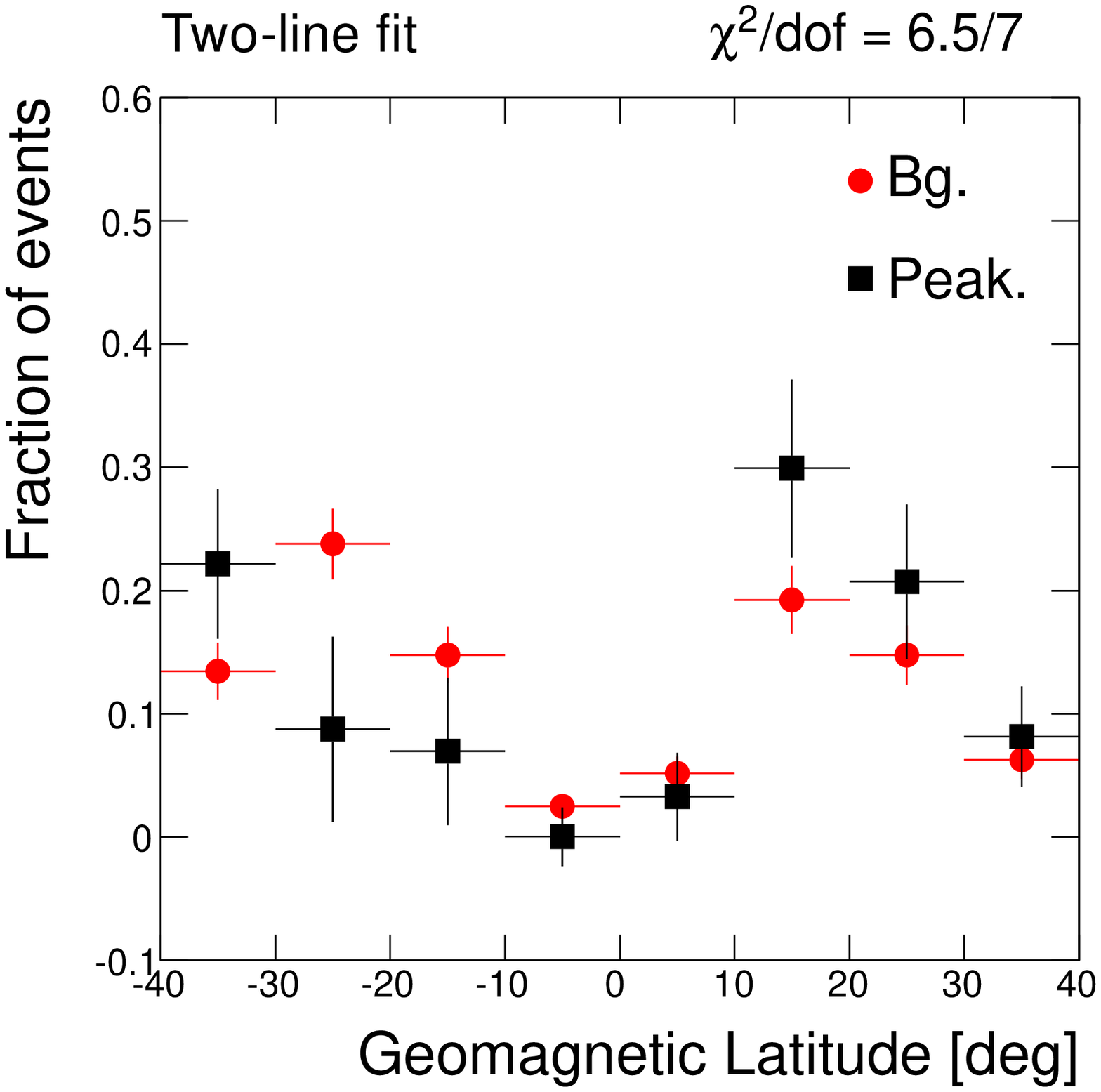}
\caption{Disentangled signal and background distributions. Left,
  the distance in Earth longitude and latitude from the center of the
  South Atlantic Anomaly. Right, geomagnetic latitude of the
  spacecraft. }
\label{fig:saa2}
\end{figure}

Unfolded distributions of incidence angles are shown in
Fig.~\ref{fig:detang2}. The distributions in galactic coordinates can
be seen in Fig.~\ref{fig:coord2}. Zenith and azimuthal angle distributions are in
Fig~\ref{fig:zenith2}, and the recorded time and conversion type are
in Fig~\ref{fig:timeback2}.  Energy and direction reconstruction
quality are in Fig~\ref{fig:probs2} and the reconstructed/raw energy
ratio as well as the first layer of the tracker with a hit are shown
in Fig~\ref{fig:ratiotrk2}.  The magnetic field parameters are shown
in Fig.~\ref{fig:mag2} and the distance from the South Atlantic anomoly
and the geomagnetic latitude are shown in Fig.~\ref{fig:saa2}. In each case, we compare the
distributions quantitatively by calculating the $\chi^2/$dof between
the peak and background distributions, shown in Table~\ref{tab:stat}.

\begin{table}
\caption{Summary of consistency between background and peak
  distributions for each of the considered instrumental variables,
  expressed as the $\chi^2$ per degree of freedom.}
\label{tab:stat}
\begin{tabular}{lrr}
\hline\hline
Variable & Single-line & Double-line \\
 & $\chi^2/$dof &  $\chi^2/$dof\\
\hline
$\cos(\theta)$ & 8.9/5 & 10.7/5\\
Detector Azmith & 4.4/7 & 7.9/7 \\
Zenith Angle & 4.3/7 & 10.2/7 \\
Earth Azimuth & 1.1/4 & 2.5/4\\
Mission Time & 1.6/7 & 2.7/7 \\
Conversion Type & 0.0/1 & 0.0/1\\
Prob correct energy & 6.8/7 & 12.1/7 \\
Prob correct dir & 2.7/7 & 6.0/7\\
Reco/Raw energy & 11.9/6 & 11.4/6\\
First tracker hit & 2.4/7 & 3.6/7 \\
McIlwain $B$  & 11.9/7 & 11.9/7 \\
McIlwain $L$  & 2.5/7 & 3.8/7 \\
Distance from SA Anomaly & 1.4/6 & 1.9/6\\
Geomagnetic Latitude & 2.6/7 &  6.5/7 \\
\hline\hline
\end{tabular}
\end{table}

\section{Sensitivity}

The number of events in the observed peak is not large, which makes
the task of identifying a potential instrumental feature difficult.
Before we can draw conclusions about the distributions above, we must
understand whether we would expect to see a feature given the current
statistics.

\begin{figure}
\includegraphics[width=1.8in]{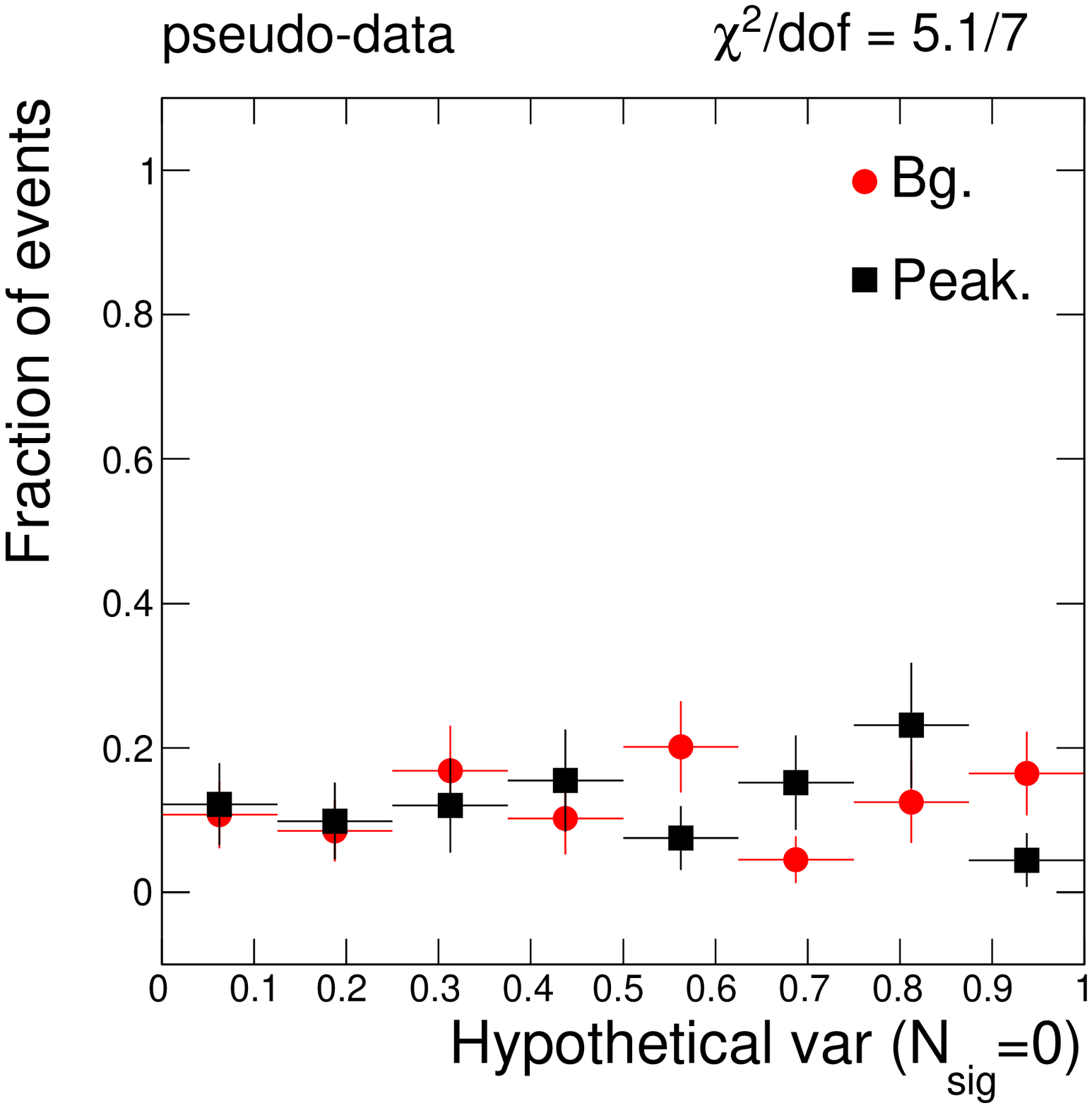}\\
\includegraphics[width=1.8in]{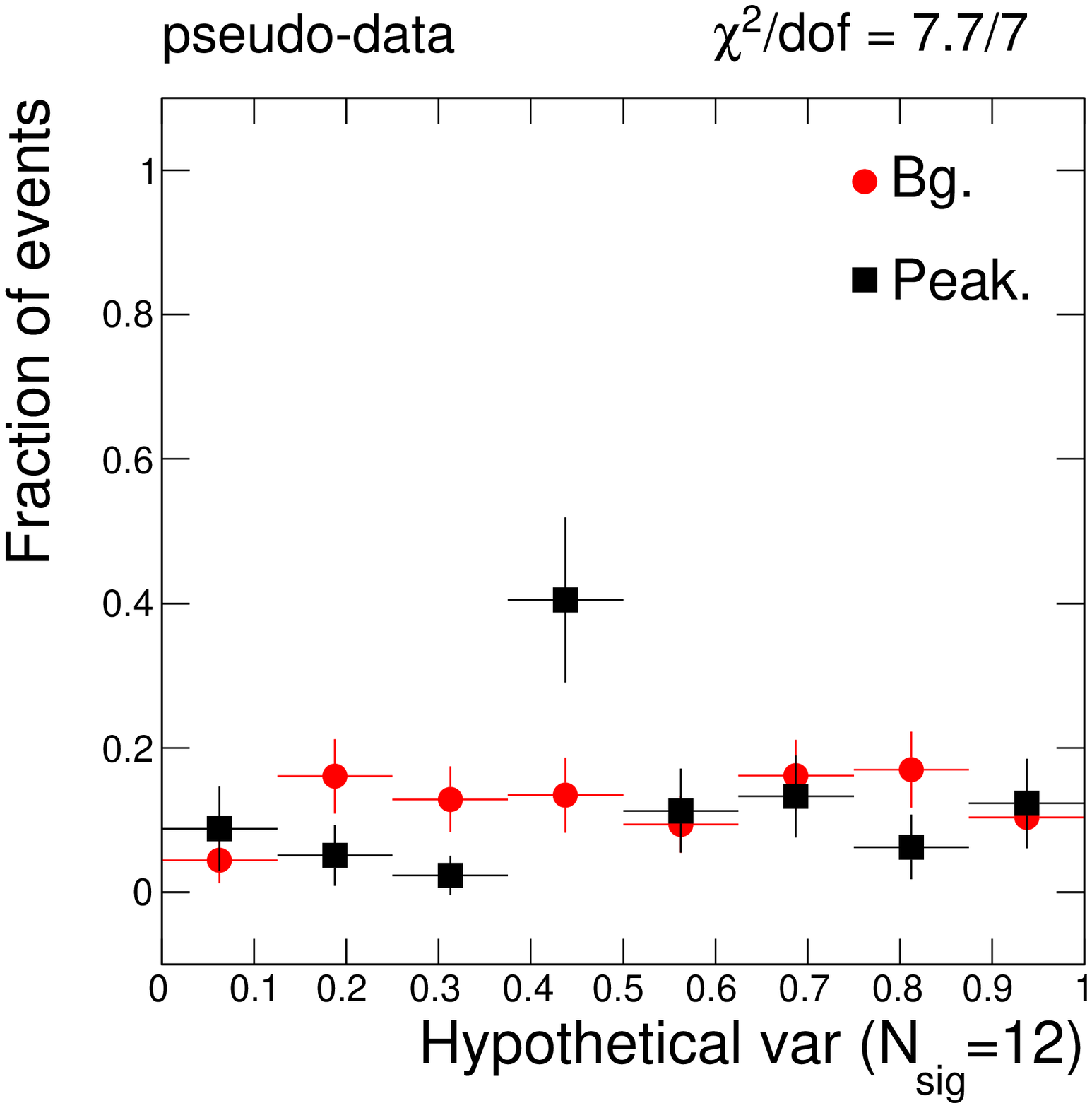}\\
\includegraphics[width=1.8in]{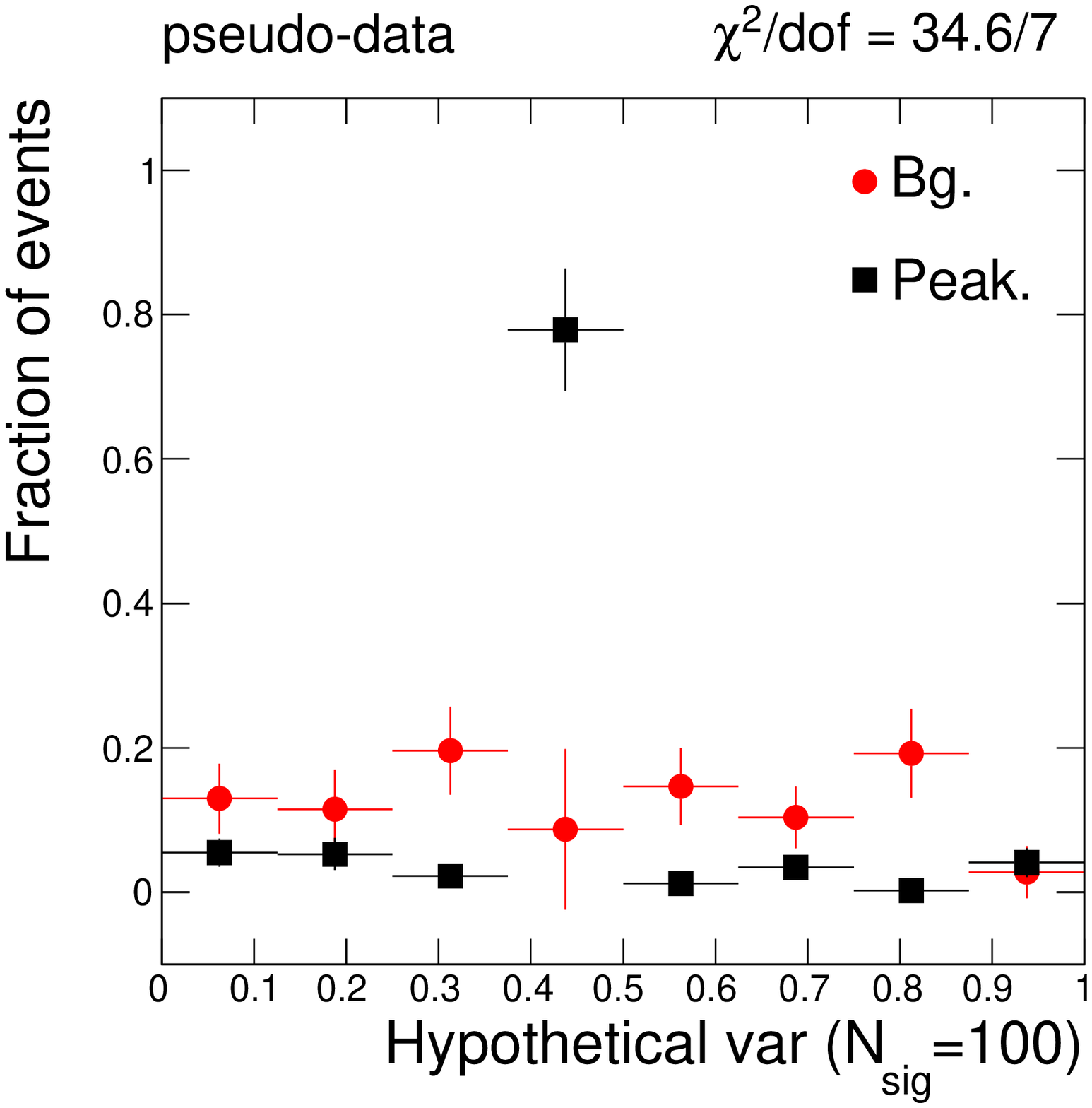}
\caption{Disentangled signal and background distributions in simulated experiments
  hypothetical variable in which the background is uniform and the
  signal is a delta function at $x=0.45$, for varying amounts of
  signal. Top, no signal events; center, 12 signal events, the
  approximate number seen in the Fermi-LAT data; bottom, 100 signal events.}
\label{fig:sens}
\end{figure}

To probe this question, we perform simulated experiments using a
hypothetical variable in which the background is uniformly distributed
between 0 and 1 and the signal peak is a delta function at 0.45; this
represents an optimistic scenario in which the entire signal is tightly clustered.
Figure~\ref{fig:sens} shows representative individual example experiments with either
zero, 12 or 100 signal events.  If the signal statistics were very large
($N_{\textrm{sig}}=100$ events), such a strong feature would be
observable both as a discrepant single bin and a $\chi^2/$d.o.f. with
low probability, $P(\chi^2/$d.o.f.$=34.6/7) = 10^{-5}$. In the
current statistics ($N_{\textrm{sig}}\approx 12$ events), the feature
would be noticeable in a single bin, but the  $\chi^2/$d.o.f., which
analyzes the global consistency of the two distributions, would be
reasonable, $P(\chi^2/$d.o.f.$=7.7/7) = 0.36$.

In the instrumental features we study here, this scenario may be
overly optimistic -- a real instrumental feature may
appear as a more subtle difference between the two distributions. It may also be pessimistic, as the signal feature could appear
where the background is suppressed, whereas in the hypothetical
variable the background is uniform.  However, the simulated
experiments suggest that even if there were a true strong instrumental
disagreement between the signal-like and background-like photons, we
may identify one or two discrepant bins, but 
are unlikely to find a $\chi^2$/dof with convincingly small probability.  This
emphasizes our earlier point, that an observed discrepancy in the
distribution of signal-like and background-like photons should serve
as a clue for further instrumental studies, rather than conclusive
evidence for or against an instrumental explanation.

As a positive control, we can examine the galactic longitude. The
feature at $E_\gamma=130$ GeV has been previoulsy localized to $l=-1.5^\circ$~\cite{finksu},
which is consistent with what we observe in Fig~\ref{fig:coord1}.
While the individual bin near $l=-1.5^\circ$ shows a large 
discrepancy between signal-like and background-like photons, the
global agreement of the distributions in longitude is reasonable,  with
a $p$-value of 0.3.

\section{Discussion}

Examining the unfolded distributions, there are several bins which
show suggestive but inconclusive discrepancies.  The distribution of
cos$(\theta)$ (Fig.~\ref{fig:detang1}) shows a bin with a large
fraction of the signal events near cos$(\theta)=0.7$. This is unlikely
to cause a feature in the energy spectrum, though the resolution
depends on cos$(\theta)$, as the cluster occurs at the median value
rather than at either extreme. The overall consistency is reasonable,
$P(\chi^2/$d.o.f.$=8.9/5)=0.11$, though see the sensitivity discussion
above.  Similarly, there is a single discrepant bin in the McIlwain
$B$ parameter at 1.65 Guass (Fig.~\ref{fig:mag1}).  These may be useful clues for further
instrumental studies.

\section{Conclusions}

We have performed an initial study of the instrumental characteristics of
events from the feature at $E_\gamma=130$ GeV observed in the
Fermi-LAT data.   

In the instrumental variables available in the
public data distribution, we find no conclusive difference in
characteristics between peak photons and background photons, see
Table~\ref{tab:stat}.  There are several suggestive
discrepancies, near cos$(\theta)=0.7$ or  McIlwain
$B$ parameter of 1.65 Guass which deserve further study by
instrumental experts.

There are several additional instrumental variables which
should be examined, such as the incident position on the face of the
LAT, but are not available in the public data.

 If a striking feature had appeared -- such as a clustering of the peak
photons at a given time or near a specific angle of incidence -- it
would have pointed to an instrumental issue.  The statistics of the
sample are too poor to draw strong conclusions, but the lack of a very
clear features makes an instrumental explanation somewhat less likely.

\acknowledgements
\section{Acknowledgements}

DW acknowledges  contributions, explanations
and useful discussions with Eric Albin which are clearly deserving of
authorship, comments from Aaron Pierce, Paddy Fox and Tim Tait, insightful statistical comments from Kyle Cranmer, and technical support from Mariangela Lisanti and Tracy Slatyer.
DW is supported by grants from the Department of Energy
Office of Science and by the Alfred P. Sloan Foundation. DW is
grateful to the Aspen Center for Physics, where this work was
performed and supported by NSF grant no. 1066293.


\begin{thebibliography}{99}


\bibitem{Abdo:2010nc}
  A.~A.~Abdo {\it et al.}  [ The Fermi-LAT Collaboration],
  Phys.\ Rev.\ Lett.\  {\bf 104}, 091302 (2010)
  arXiv:1001.4836 

\bibitem{Fermi:2012}
  M.~Ackermann {\it et al.}  [Fermi-LAT Collaboration], (2012),
  arXiv:1205.2739




\bibitem{Bringmann:2012vr}
  T.~Bringmann, X.~Huang, A.~Ibarra, S.~Vogl and C.~Weniger, (2012),
  arXiv:1203.1312 

\bibitem{Weniger:2012tx}
  C.~Weniger, (2012),
  arXiv:1204.2797 

\bibitem{Tempel:2012ey}
  E.~Tempel, A.~Hektor and M.~Raidal, (2012),
  arXiv:1205.1045 
   
\bibitem{finksu}
M.~Su and D.~P.~Finkbeiner, (2012),
  arXiv:1206.1616 

\bibitem{wacker}
T.~Cohen, M.~Lisanti, T.~R.~Slatyer and J.~G.~Wacker, (2012),
  arXiv:1207.0800 

\bibitem{splots}
  M.~Pivk and F.~R.~Le Diberder,
  Nucl.\ Instrum.\ Meth.\ A {\bf 555}, 356 (2005)
  physics/0402083 
 
\bibitem{astrocite} {\sc inspire} search for ``refersto:recid:644725''
  yields no items posted to astro-ph.

\bibitem{qual} Pass7,  \texttt{ultraclean} class, quality
  requirements: {\texttt{DATA\_QUAL$=1$ \&\& LAT\_CONFIG$=$1 \&\&
      ABS(ROCK\_ANGLE)$\le$52 \&\& ZENITH<100}} and a good time
  interval (via \texttt{gtmktime}).

\bibitem{blah:2012kca} 
  [Fermi-LAT Collaboration],
  arXiv:1206.1896 [astro-ph.IM].


\bibitem{fermiedisp}
\url{http://fermi.gsfc.nasa.gov/ssc/data/analysis/documentation/Cicerone/Cicerone_LAT_IRFs/IRF_E_dispersion.html}

\bibitem{fermidefs}
\url{http://fermi.gsfc.nasa.gov/ssc/data/analysis/documentation/Cicerone/Cicerone_Data/LAT_Data_Columns.html}

\bibitem{mcilwain} McIlwain, C. E.,  J. Geophys. Res. 66, pp. 3681-3691 (1961).

\bibitem{Atwood:2009ez} 
  W.~B.~Atwood {\it et al.}  [LAT Collaboration],
  Astrophys.\ J.\  {\bf 697}, 1071 (2009)
  [arXiv:0902.1089 [astro-ph.IM]].

\bibitem{Abdo:2009gy} 
  A.~A.~Abdo {\it et al.}  [Fermi LAT Collaboration],
  Astropart.\ Phys.\  {\bf 32}, 193 (2009)
  [arXiv:0904.2226 [astro-ph.IM]].


\bibitem{twolines}
  A.~Rajaraman, T.~M.~P.~Tait and D.~Whiteson, JCAP, accepted (2012),
  arXiv:1205.4723. 



\end{thebibliography}
\end{document}